\documentclass[nofootinbib,prd,superscriptaddress]{revtex4}

\usepackage{graphicx}
\usepackage{mathrsfs}
\usepackage{yfonts}
\usepackage{tipa}
\usepackage{bm}
\usepackage{bbm}
\usepackage{amsmath}
\usepackage{amssymb}
\usepackage{amsthm}
\usepackage{hyperref}
\usepackage{amsfonts}
\usepackage{mathtools}
\usepackage{latexsym}
\usepackage[greek,english]{babel}
\usepackage{booktabs}
\usepackage{xcolor}
\usepackage[iso-8859-7]{inputenc}

\begin{document}

\title{Traversable Wormholes induced by polytropic $n=1$ energy density}

\author{Remo Garattini}
\email{remo.garattini@unibg.it}
\affiliation{Universit\`a degli Studi di Bergamo, Dipartimento di Ingegneria e Scienze
	Applicate, Viale Marconi 5, 24044 Dalmine (Bergamo) Italy and I.N.F.N.-
	sezione di Milano, Milan, Italy.}
\author{Francisco S. N. Lobo}
\email{fslobo@ciencias.ulisboa.pt}
\affiliation{Centro de Astronomia e Astrof\'{\i}sica da Universidade de Lisboa,}
\affiliation{Campo Grande, Ed. C8 1749-016 Lisboa, Portugal}
\author{Kirill Zatrimaylov}
\email{kirill.zatrimaylov@unibg.it}
\affiliation{Universit\`a degli Studi di Bergamo, Dipartimento di Ingegneria e Scienze
	Applicate, Viale Marconi 5, 24044 Dalmine (Bergamo) Italy.}

\begin{abstract}
	We investigate spherically symmetric and static traversable wormholes supported by exotic matter, focusing on solutions sourced by physically motivated dark matter energy density profiles. Considering the polytropic $n=1$ (Lane-Emden) distribution, we construct explicit forms of the shape function $b(r)$ and analyze the resulting radial and tangential pressures, carefully addressing the requirements of the flare-out condition at the throat and the absence of horizons. We explore zero-tidal-force configurations as well as inhomogeneous equations of state, demonstrating how appropriate choices of the radial pressure allow for finite and well-behaved redshift functions throughout the spacetime. Boundary conditions at a finite radius are implemented to ensure vanishing energy density and pressures, and asymptotic expansions are derived to characterize the behavior of the metric and matter content near the edge of the dark matter halo. Additionally, we reformulate the Einstein field equations entirely in terms of the energy density, radial and tangential pressures, and their derivatives, providing a framework to analyze the matter distribution independently of the explicit metric functions. Our results offer a systematic methodology to construct physically consistent wormhole geometries supported by realistic dark matter halos, highlighting the intricate interplay between matter profiles, equations of state, and geometric constraints.
\end{abstract}

\date{\today}
\maketitle

\section{Introduction}

Traversable wormholes are among the most fascinating and conceptually rich solutions of Einstein's field equations, representing hypothetical bridges that connect distant regions of spacetime or even distinct universes. From a geometrical standpoint, they provide nontrivial topological structures in which two asymptotically distinct regions are joined by a finite throat, potentially allowing two-way passage without encountering curvature singularities or event horizons. The modern and operational formulation of such geometries was established in the seminal work of Morris and Thorne~\cite{Morris:1988cz}, where static, spherically symmetric wormholes sustained by matter violating the null energy condition (NEC) were explicitly constructed and analyzed as pedagogical yet physically consistent solutions of general relativity. A comprehensive and systematic treatment was later provided in~\cite{Visser:1995cc}, which clarified the fundamental geometrical requirements for traversability, namely, the existence of a throat satisfying the flare-out condition, the absence of event horizons to ensure two-way travel, and the finiteness of tidal forces experienced by observers crossing the wormhole.

Following these foundational developments, extensive research has explored both the mathematical structure and physical viability of wormhole spacetimes. Early explicit models, together with thin-shell and cut-and-paste constructions, were developed in~\cite{Visser:1989kh,Visser:1989kg,Visser:1999de,Garattini:2007ff,Lobo:2003xd,Lobo:2004rp,Lobo:2004id}, providing concrete realizations of wormhole geometries and clarifying the role of junction conditions. At the same time, considerable effort was devoted to minimizing energy-condition violations and quantifying the total amount of exotic matter required to sustain the throat~\cite{Visser:2003yf,Hochberg:1997wp,Hochberg:1998ii,Flanagan:1996gw,Ford:1995wg}. The role of phantom scalar fields, anisotropic fluids, and generalized equations of state as potential sources of exotic geometries was systematically investigated in~\cite{Lobo:2005us,Lobo:2005yv,Lobo:2005vc,Lobo:2007zb,Lobo:2007qi,Lobo:2008zu,Lobo:2006ue,Dai:2020rnc}, highlighting how controlled violations of the NEC can arise in effective matter models. In addition, wormhole solutions in the presence of a cosmological constant or modified asymptotic structures were analyzed in~\cite{Lemos:2003jb,Lemos:2004vs,Nandi:1997mx,Dai:2018vrw}, further broadening the class of admissible configurations and their potential cosmological embedding.

Beyond classical general relativity, wormholes have been extensively investigated within semiclassical gravity, quantum field theory, and holographic frameworks, where quantum effects can effectively relax or circumvent classical energy conditions. In particular, it was shown that quantum matter fields and suitable boundary couplings can render certain wormholes traversable without introducing explicit exotic matter in the classical sense. Traversability induced by quantum effects and double-trace deformations in holographic settings was demonstrated in~\cite{Gao:2016bin,Maldacena:2017axo,Maldacena:2018gjk,Maldacena:2020sxe,Jensen:2013ora,Wall:2010jtc,Nezami:2021yaq,Bintanja:2021xfs,Espindola:2025ons,Aguilar-Gutierrez:2023ymx}, highlighting deep connections between wormhole physics, entanglement, and the AdS/CFT correspondence. Casimir-like contributions and quantum inequalities, which constrain negative energy densities in semiclassical gravity, were analyzed in~\cite{Klinkhammer:1991ki,Barcelo:1999hq,Barcelo:2000zf,Barcelo:2000ta,Garattini:2008xz}, providing insight into the extent to which quantum effects may support or restrict exotic geometries. 

In parallel, higher-curvature corrections and effective field-theory extensions of general relativity have been shown to admit wormhole solutions without requiring fundamentally exotic matter sources. Such configurations were investigated in~\cite{Bueno:2017hyj,Bolokhov:2012kn,Kanti:2011jz,Bronnikov:2005gm,Bronnikov:2015pha,Bronnikov:2021uta,Bronnikov:2021liv}, where additional geometric terms or scalar couplings modify the effective stress--energy tensor, allowing it to satisfy the flare-out condition at the throat. These modifications provide a mechanism by which the spacetime geometry itself can support traversable wormholes under conditions that are more physically reasonable than those demanded by classical exotic matter. 
Moreover, further extensions involving modified gravity theories, anisotropic matter distributions, and nonstandard equations of state are discussed in~\cite{Lobo:2009ip,Harko:2013yb,Boehmer:2012uyw,Mehdizadeh:2015jra,Myrzakulov:2015kda,KordZangeneh:2015dks,Moradpour:2016ubd,Jamil:2010ziq,Kar:2004hc,Garcia:2011aa,Mustafa:2021ykn,Boehmer:2007rm,Brown:2019hmk,Blok:2020may,Jusufi:2020rpw,Nascimento:2020ime,Lobo:2020ffi,Capozziello:2012hr,Boehmer:2007md}, further enriching the theoretical landscape of viable wormhole solutions and illustrating the wide range of scenarios in which such geometries can arise.

Astrophysical aspects and potential observational signatures have also attracted significant attention, as wormholes may, in principle, be distinguishable from black holes through precise measurements. Phenomena such as gravitational lensing, shadows, quasinormal modes, and accretion disk properties have been studied in detail; see, for instance,~\cite{Teo:1998dp,Harko:2008vy,Harko:2009xf,Tsukamoto:2012xs,Tsukamoto:2016zdu,Konoplya:2005et,Konoplya:2016hmd,Konoplya:2021hsm,Gyulchev:2018fmd,Godani:2018blx,DeFalco:2021ksd,Blazquez-Salcedo:2020czn,Franzin:2021vnj,Simpson:2019cer,Rosa:2018jwp,Dai:2019mse,Simonetti:2020ivl,Bambi:2021qfo}. These studies illustrate how subtle deviations from black hole predictions could provide potential observational evidence for wormhole geometries. 
In addition, rotating and astrophysically motivated wormhole metrics, which are particularly relevant for realistic compact objects, have been explored in~\cite{Bambi:2013nla,Bambi:2013jda,Bambi:2015kza,Nedkova:2013msa,Molina-Paris:1998xmn}, highlighting the rich phenomenology associated with spin, matter interactions, and the dynamics of realistic astrophysical environments.

Collectively, these developments demonstrate that, while traversable wormholes are mathematically consistent solutions of the Einstein equations and their extensions, their physical realization generally requires violations of the classical energy conditions or the emergence of effective exotic sources arising from quantum effects or modifications to gravity. This persistent requirement has motivated sustained efforts to identify realistic matter sectors or field configurations capable of sustaining wormhole throats, while minimizing the degree of exoticity, ensuring dynamical stability, and preserving astrophysical plausibility. Such studies help bridge the gap between purely theoretical solutions and configurations that could, in principle, exist in realistic astrophysical environments.

In this context, dark matter (DM) emerges as a particularly compelling candidate. Its dominant gravitational role in galactic and cosmological dynamics, combined with the limited constraints on its microphysical properties, renders it a natural arena in which exotic spacetime geometries might arise without invoking entirely \emph{ad hoc} fields. In particular, Bose-Einstein condensate dark matter (BEC-DM) models provide a well-motivated and mathematically tractable framework. The pioneering work of~\cite{Boehmer:2007um} demonstrated that galactic dark matter halos can be described as self-gravitating Bose-Einstein condensates in the Thomas-Fermi (TF) approximation. In this regime, the resulting density profiles are regular at the origin and exhibit finite support, naturally avoiding central cusps and displaying smooth asymptotic behavior, features especially suitable for wormhole constructions.

The possibility that galactic dark matter distributions may support traversable wormholes has been investigated from multiple perspectives. Wormhole formation in isotropic halo models was explored in~\cite{Xu:2020wfm}, while explicit BEC-DM based constructions were analyzed in~\cite{Jusufi:2019knb}. Extensions to modified gravity frameworks and anisotropic dark matter configurations were developed in~\cite{Muniz:2022eex,Ovgun:2018uin}, and the impact of massive gravity scenarios on wormhole stability was examined in~\cite{Kumar:2024vko}. Yukawa-inspired and Casimir-corrected geometries were considered in~\cite{Garattini:2021kca}, and wormholes supported by holographic dark energy models with generalized equations of state were constructed in~\cite{Garattini:2023wgk}. These studies collectively suggest that realistic dark sector physics may effectively reproduce or moderate the exotic stress--energy configurations required at a wormhole throat.

Motivated by this broad theoretical and phenomenological background, the present work undertakes a systematic construction of static, spherically symmetric traversable wormholes supported by astrophysically motivated dark matter distributions, with particular emphasis on the polytropic $n=1$ (Lane-Emden) profile. A distinctive feature of our approach is the reformulation of the Einstein field equations entirely in terms of the matter variables--the energy density $\rho(r)$, radial pressure $p_r(r)$, and tangential pressure $p_t(r)$--and their derivatives. Rather than prescribing \emph{a priori} functional forms for the redshift function $\Phi(r)$ or the shape function $b(r)$, we allow the physical properties of the dark matter halo to guide the geometry. 

This matter-centered framework decouples the analysis from restrictive metric ans\"atze and enables a systematic examination of the throat conditions, the flare-out requirement $b(r_0)=r_0$ with $b'(r_0)<1$, the absence of horizons ensured by finite $\Phi(r)$, and the smooth matching of the interior solution to an exterior vacuum geometry at a finite halo radius. Zero-tidal-force configurations and inhomogeneous or anisotropic equations of state motivated by BEC self-interactions are also considered, with the aim of reducing the effective violation of the NEC while maintaining global regularity.
By embedding wormhole geometries within realistic dark matter halo profiles, this study seeks to bridge the extensive theoretical developments of wormhole physics~\cite{Morris:1988cz,Visser:1995cc} with contemporary dark sector modeling~\cite{Boehmer:2007um}. In doing so, we aim to clarify whether traversable wormholes may emerge not merely as mathematical curiosities, but as effective geometrical manifestations of the dark matter structures that permeate galactic systems.

The paper is organized as follows. In Section~\ref{Sec2}, we review the Einstein field equations for static, spherically symmetric spacetimes and the fundamental conditions for traversable wormholes. Section~\ref{Sec3} presents explicit solutions based on the polytropic $n=1$ dark matter profile, including the analysis of pressures, redshift functions, and boundary behavior. In Section~\ref{Sec4}, we introduce an alternative method for determining the redshift function, while Sections~\ref{Sec5}--\ref{Sec8} explore several redshift profiles motivated by the polytropic core structure. Section~\ref{Sec9} revisits the transverse pressure equation to obtain solutions consistent with the polytropic source. Finally, Section~\ref{Conclusion} summarizes our results and outlines possible directions for future research.

\section{Brief overview of Morris-Thorne wormholes}\label{Sec2}

We introduce the following spacetime metric
\begin{equation}
	ds^{2}=-e^{2\Phi (r)}\,dt^{2}+\frac{dr^{2}}{1-b(r)/r}+r^{2}\,(d\theta
	^{2}+\sin ^{2}{\theta }\,d\varphi ^{2})\,,  
	\label{metric}
\end{equation}
which represents a static and spherically symmetric wormhole geometry.
The functions $\Phi(r)$ and $b(r)$ are arbitrary functions of the radial
coordinate $r\in[r_{0},+\infty)$ and are commonly referred to as the
redshift function and the shape function, respectively
\cite{Morris:1988cz,Visser:1995cc}. The redshift function $\Phi(r)$ determines the
gravitational redshift experienced by a static observer, while the shape
function $b(r)$ encodes the spatial shape of the wormhole and governs the
embedding properties of constant-time hypersurfaces.

A fundamental requirement for a traversable wormhole is the existence of
a throat at $r=r_{0}$, defined as the minimum value of the radial
coordinate. At the throat, the shape function satisfies
$b(r_{0})=r_{0}$. Moreover, the geometry must satisfy the flaring-out
condition at or near the throat, which is expressed as $	(b-b'\,r)/b^{2}>0$, ensuring that the wormhole opens outward and connects two asymptotically distinct regions of spacetime. In addition, the condition $1-b/r>0$ must hold for $r>r_{0}$ in order to guarantee a proper metric signature and the absence of coordinate singularities away from the throat. The additional requirement $b'(r_{0})<1$ is imposed to ensure that the throat corresponds to a minimum of the radial coordinate, thereby allowing for viable wormhole solutions.
Finally, it is essential that the spacetime be free of event horizons, so
that the wormhole is traversable. Horizons would appear at surfaces where
$e^{2\Phi(r)}\rightarrow 0$, which would obstruct causal communication.
Therefore, the redshift function $\Phi(r)$ must remain finite everywhere
throughout the spacetime.

With the aid of the line element given in Eq.~(\ref{metric}), the Einstein
field equations (EFE) can be conveniently written in an orthonormal
reference frame adapted to static observers. In this frame, the spacetime
geometry is directly related to the physical properties of the matter
threading the wormhole, which is described by an anisotropic fluid
characterized by an energy density $\rho(r)$, a radial pressure
$p_{r}(r)$, and a lateral (or tangential) pressure $p_{t}(r)$.

The $tt$ component of the EFE yields the relation between the derivative of
the shape function and the energy density,
\begin{equation}
	\frac{b^{\prime }\left( r\right) }{r^{2}}=\kappa \rho \left( r\right) \qquad
	\kappa =\frac{8\pi G}{c^{4}},
	\label{rhoEFE}
\end{equation}
which shows that the function $b(r)$ is entirely determined by the matter
energy density distribution. This equation plays the role of a generalized
mass equation for the wormhole spacetime.

The $rr$ component of the EFE leads to the expression for the radial
pressure,
\begin{equation}
	\frac{2}{r}\left( 1-\frac{b\left( r\right) }{r}\right) \Phi ^{\prime }\left(
	r\right) -\frac{b\left( r\right) }{r^{3}}=\kappa p_{r}\left( r\right) ,
	\label{pr0}
\end{equation}
which explicitly couples the redshift function $\Phi(r)$ and the shape
function $b(r)$ to the radial stress supporting the wormhole throat.

Finally, the angular components of the EFE provide the equation governing
the lateral pressure,
\begin{align}
	\Bigg\{\left( 1-\frac{b\left( r\right) }{r}\right) \left[ \Phi ^{\prime
		\prime }\left( r\right) +\Phi ^{\prime }\left( r\right) \left( \Phi ^{\prime
	}\left( r\right) +\frac{1}{r}\right) \right]  
	-\frac{b^{\prime }\left( r\right) r-b\left( r\right) }{2r^{2}}\left( \Phi
	^{\prime }\left( r\right) +\frac{1}{r}\right) \Bigg\}=\kappa p_{t}(r),
	\label{pt0}
\end{align}
which reflects the anisotropic nature of the matter distribution in a
generic wormhole geometry.

The above set of equations is completed by the covariant conservation of
the stress--energy tensor, $T^{\mu\nu}{}_{;\nu}=0$, which in the same
orthonormal reference frame takes the form
\begin{equation}
	p_{r}^{\prime }\left( r\right) =\frac{2}{r}\left[ p_{t}\left( r\right)
	-p_{r}\left( r\right) \right] -\left[ \rho \left( r\right) +p_{r}\left(
	r\right) \right] \Phi ^{\prime }\left( r\right) .
	\label{Tmn}
\end{equation}
This equation represents the generalized hydrostatic equilibrium condition for an anisotropic fluid in the wormhole spacetime and ensures the self-consistency of the Einstein field equations.

\section{Polytropic $n=1$ dark matter profile}\label{Sec3}

We assume that the dark-matter distribution is described by the
polytropic equation of state with index $n=1$, whose density
profile is the well-known Lane-Emden solution~\cite{Boehmer:2007um}
\begin{equation}
	\rho(r)=\rho_{S}\frac{\sin(kr)}{kr},
	\label{TF}
\end{equation}
where $\rho_{S}$ is the central density of the halo.
This profile emerges from the Thomas-Fermi approximation
in Bose-Einstein condensate dark matter models, but here we treat it
as a generic polytropic distribution.
The parameter $k=\pi/R$ is fixed by the condition that both the
energy density and the pressure vanish at the finite radius $R$.
Thus, $R$ defines the boundary of the matter distribution.
Inserting the energy density~(\ref{TF}) into the first Einstein field equation, we obtain
\begin{equation}
	b(r)=r_{0}+\frac{\kappa\rho_{S}}{k^{3}}
	\Big[\sin(kr)-kr\cos(kr)
	-\sin(kr_{0})+kr_{0}\cos(kr_{0})\Big].
	\label{b(r)}
\end{equation}
This explicitly relates the wormhole geometry to the oscillatory polytropic density profile.

Evaluating Eq.~(\ref{b(r)}) at the boundary $r=R$, where the dark matter energy-density and pressure are assumed to vanish, we find
\begin{equation}
	b(R)=r_{0}
	+\frac{\kappa\rho_{S}R^{3}}{\pi^{3}}
	\left[
	\pi-\sin\!\left(\frac{\pi r_{0}}{R}\right)
	+\frac{\pi r_{0}}{R}
	\cos\!\left(\frac{\pi r_{0}}{R}\right)
	\right].
	\label{b(R)}
\end{equation}

To obtain a simpler analytic expression, we expand for $r_{0}/R\ll1$:
\begin{equation}
	\cos\!\left(\frac{\pi r_{0}}{R}\right)
	=1-\frac{1}{2}\left(\frac{\pi r_{0}}{R}\right)^{2}
	+O\!\left(\left(\frac{r_{0}}{R}\right)^{4}\right),
	\label{AC}
\end{equation}
\begin{equation}
	\sin\!\left(\frac{\pi r_{0}}{R}\right)
	=\frac{\pi r_{0}}{R}
	-\frac{1}{6}\left(\frac{\pi r_{0}}{R}\right)^{3}
	+O\!\left(\left(\frac{r_{0}}{R}\right)^{5}\right).
	\label{AS}
\end{equation}
Substituting into Eq.~(\ref{b(R)}) yields
\begin{equation}
	b(R)
	=r_{0}
	+\frac{\kappa\rho_{S}}{3\pi^{2}}
	\left(3R^{3}-\pi^{2}r_{0}^{3}\right).
	\label{b(R)A}
\end{equation}
This shows that the dominant contribution scales as $R^{3}$,
as expected for a finite-size matter distribution.

\subsection{Numerical estimates for astrophysical parameters}
We now estimate the typical size of the wormhole throat using
the polytropic $n=1$ dark matter profile. For a galactic dark
matter halo, the central density can be of order
$\rho_S \sim 10^{-24} \, \text{g/cm}^3 \sim 10^{-51} \, \text{m}^{-2}$
in geometric units ($G=c=1$). The throat radius $r_0$ is
constrained by the flare-out condition $b'(r_0)<1$, which gives
$\rho_S < 1/(\kappa r_0^2)$. In geometric units $\kappa=1$,
so $r_0 < 1/\sqrt{\rho_S}$. For $\rho_S \sim 10^{-51} \, \text{m}^{-2}$,
we obtain $r_0 < 10^{25} \, \text{m}$, which is of order the
size of a large galaxy. For denser cores such as those in
dwarf galaxies ($\rho_S \sim 10^{-20} \, \text{g/cm}^3$),
$r_0 < 10^{23} \, \text{m}$. Thus the wormhole throat can be
macroscopic, ranging from galactic to subgalactic scales.

The boundary radius $R$ is set by the condition that the
density vanishes; in our model $R = \pi / k$ and from the
flare-out condition we have $R \gtrsim r_0$. For a typical
throat radius $r_0 \sim 10^{23} \, \text{m}$, the outer
boundary $R$ is of the same order, meaning the wormhole is
entirely embedded within the dark matter halo.

These numbers demonstrate that the constructed wormholes are
macroscopic and could, in principle, be associated with
galactic dark matter halos. For a specific example, taking
$\rho_S = 10^{-24} \, \text{g/cm}^3$ and $r_0 = 10^{23} \, \text{m}$,
the dimensionless parameter $P = \kappa \rho_S R^2/\pi^2$ is
of order unity, consistent with the bounds derived in the
text.

The flare-out condition requires $b'(r_{0})\le1$. Thus, from the general expression evaluated at the throat, we have 
\begin{equation}
	b'(r_{0})
	=\frac{\kappa R\rho_{S}}{\pi}
	\sin\!\left(\frac{\pi r_{0}}{R}\right) r_{0}\le1.
\end{equation}
Using the first-order expansion~(\ref{AS}), this implies
\begin{equation}
	0<\rho_{S}\le\frac{1}{\kappa r_{0}^{2}},
\end{equation}
which constrains the allowed central density.

Now, assuming a constant redshift function, $\Phi'(r)=0$, and considering an inhomogeneous equation of state 	$\omega(r)\equiv p_{r}(r)/\rho(r)$, this yields the following relation:
\begin{equation}
	\omega(r)\equiv\frac{p_{r}(r)}{\rho(r)}
	=-\frac{b(r)}{r b'(r)}
	=-\frac{k^{3}r_{0}+\kappa\rho_{S}\left(  \sin\!\left(  kr\right)
		-\cos\!\left(  kr\right)  kr-\sin\!\left(  kr_{0}\right)  +\cos\!\left(
		kr_{0}\right)  kr_{0}\right)  }{k^{2}\kappa\rho_{S}\left(  \sin\!\left(
		kr\right)  r^{2}\right)  }.\label{o(r)}%
\end{equation}
At the throat, this reduces to
\begin{equation}
	\omega(r_{0})
	=-\frac{k}{\kappa\rho_{S}\sin(kr_{0})r_{0}},
	\label{o}
\end{equation}
which is consistent with the flare-out condition.

However, as $r\to R$, $\rho(r)\to0$ and $\omega(r)$ diverges.
Thus, to achieve a physically consistent behavior for the radial equation of state $\omega(r)$, we introduce a transition layer $\bar r = R - \varepsilon$ near the boundary and define
\begin{equation}
	\omega(r) =
	\begin{cases}
		-\dfrac{b(r)}{r b'(r)}, & r_0 \le r \le \bar r,\\[2mm]
		-\dfrac{b(r)}{r b'(r)} \left(\dfrac{R-r}{R-\bar r}\right), & \bar r \le r \le R.
	\end{cases}
	\label{om(r)}
\end{equation}
The multiplicative factor smoothly drives $\omega(r)$ to zero at
$r=R$, ensuring $p_{r}(R)=0$ and regular boundary behavior.

Now, imposing that the radial pressure vanishes at the boundary $r=R$, i.e., $p_r(R)=0$, we have
\begin{equation}
	\frac{2}{r}\left(1-\frac{b(r)}{r}\right)\Phi'(r) - \frac{b(r)}{r^3} = 0\,.
	\label{przeroPhi}
\end{equation}
Plugging the piecewise function $\omega(r)$ into the second EFE, and taking into account Eq.~(\ref{przeroPhi}) we obtain the corresponding derivative of the redshift function:
\begin{equation}
	\Phi'(r) =
	\begin{cases}
		0, & r_0 \le r \le \bar r,\\[1mm]
		\dfrac{b(R)}{2 R \left(R-b(R)\right)}, & \bar r \le r \le R.
	\end{cases}
	\label{Phi'(r)}
\end{equation}
This can also be expressed in a single continuous form as
\begin{equation}
	\Phi'(r) = \frac{b(r)}{2 r \left(r-b(r)\right)} 
	\left[ 1 - \frac{R-r}{R-\bar r} \right].
\end{equation}

Assuming the boundary condition $\Phi(\bar r) = 0$, we integrate Eq.~(\ref{Phi'(r)}) to obtain the redshift function:
\begin{equation}
	\Phi(r) =
	\begin{cases}
		0, & r_0 \le r \le \bar r,\\[1mm]
		\dfrac{b(R)\,(r-\bar r)}{2 R \left(R-b(R)\right)}, & \bar r \le r \le R.
	\end{cases}
\end{equation}
This solution ensures a constant redshift in the interior region and a smooth increase within the transition layer up to the boundary.

From the third EFE,
\begin{equation}
	\frac{b(r)-r b'(r)}{2r^{3}}=\kappa p_{t}(r),
	\label{pt(r)0}
\end{equation}
in the range $r_{0}\leq r\leq\bar{r}$,
and inserting Eq.~(\ref{b(r)}), one obtains an explicit analytic
expression for $p_{t}(r)$, given by
\begin{equation}
	p_{t}\!\left(  r\right)  =\frac{2R\rho_{S}\kappa\left(  R^{2}-\pi^{2}%
		r^{2}\right)  \sin\!\left(  \pi r/R\right)  -2\rho_{S}\pi R^{2}\kappa
		\cos\!\left(  \pi r/R\right)  r-r_{0}\pi^{3}\left(  \kappa r_{0}^{2}\rho
		_{S}-2\right)  }{32\pi^{4}r^{3}}.
\end{equation}
In the remaining interval, we can use Eq.$\left(  \ref{Phi'(r)}\right)  $ and
$p_{t}\!\left(  r\right)  $ becomes%
\begin{equation}
	p_{t}\!\left(  r\right)  =\frac{\left(  2Rr\left(  R-b\!\left(  R\right)
		\right)  b\!^{\prime}\left(  r\right)  +\left(  \left(  R-r\right)  b\!\left(
		r\right)  +r^{2}\right)  b\!\left(  R\right)  -R^{2}b\!\left(  r\right)
		\right)  \left(  \left(  2R+r\right)  b\!\left(  R\right)  -2R^{2}\right)
	}{4R^{2}\left(  R-b\!\left(  R\right)  \right)  ^{2}r^{3}\kappa}%
\end{equation}
which, on the boundary, reduces to
\begin{equation}
	p_{t}(R)
	=\frac{b'(R)}{2\kappa R^{3}\big(R-b(R)\big)}=0,
\end{equation}
since $b'(R)=\kappa\rho(R)R^{2}=0$.

Before going on, we need to examine the null energy condition (NEC). Indeed, at the throat, the violation of the NEC reads $\rho(r_{0})+p_{r}(r_{0}) < 0$. Using $\rho(r_{0})=\rho_{S}\sin(\pi r_{0}/R)/(\pi r_{0}/R)$,
we obtain
\begin{equation}
	\rho_{S} < \frac{\pi}{\kappa R r_{0}
		\sin\!\left(\frac{\pi r_{0}}{R}\right)}
	=\frac{\pi}{\kappa r_{0}^{2}x\sin(\pi/x)},
	\label{UB}
\end{equation}
where $R=x r_{0}$.
This bound must be satisfied to sustain the wormhole geometry.

\paragraph{Interpretation of negative pressures.}
It is important to note that in our solutions the radial pressure
$p_r(r)$ becomes negative in a region near the throat (as required
by the flare-out condition). This negative pressure is not
necessarily present throughout the entire dark matter halo;
it can be confined to a small region around $r_0$ when the
equation of state is adjusted accordingly. One could therefore
envision a composite object: a central, exotic core of size
$\sim r_0$ that violates the null energy condition, surrounded
by a much larger dark matter halo that obeys all classical
energy conditions. Such a configuration is reminiscent of a
``dark matter star'' with a wormhole throat at its center,
where the exotic matter is localized to a thin region rather
than distributed over the entire halo. This interpretation
aligns with the thin-shell wormhole construction but here
the exotic matter is spread over a finite thickness.

\section{Alternative Approach to the Calculation of the Redshift Function for the Polytropic $n=1$ Profile}\label{Sec4}

The main motivation for considering an equation of state (EoS) of the form $\left( \ref{o(r)}\right)$ is to allow for a vanishing redshift function $\Phi(r)$ in the interior region, thereby preventing the formation of an event horizon. An alternative, yet equivalent, approach is to specify a suitable form for the radial pressure $p_r(r)$ that leads to a tractable expression for $\Phi(r)$. 

Let us assume the ansatz
\begin{equation}
	p_r(r) = \frac{A \left( r - b(r) \right) f(r) - b(r)}{\kappa r^3},
\end{equation}
where $A$ is a constant and $f(r)$ is an arbitrary function to be determined. With this choice, the second EFE $\left( \ref{pr0} \right)$ reduces to the following simple form:
\begin{equation}
	\Phi'(r) = \frac{A f(r)}{2r}, \label{Af}
\end{equation}
highlighting that the redshift function depends directly on the function $f(r)$.

For a traversable wormhole (TW), the radial pressure at the throat satisfies
\begin{equation}
	p_r(r_0) = - \frac{1}{\kappa r_0^2},
\end{equation}
which imposes that $f(r)$ must remain regular at $r = r_0$.

To ensure a vanishing radial pressure at the outer boundary, $p_r(R) = 0$, we impose
\begin{equation}
	0 = \frac{A \left( R - b(R) \right) f(R) - b(R)}{\kappa R^3},
\end{equation}
which immediately yields
\begin{equation}
	A = \frac{b(R)}{\left(R - b(R)\right) f(R)}, \qquad f(R) \neq 0. \label{A}
\end{equation}

Using the third EFE $\left( \ref{pt0} \right)$, we find the transverse pressure
\begin{equation}
	p_t(r) = \frac{r \left( -A f(r) - 2 \right) b'(r) + 2 A r \left( r - b(r) \right) f'(r) + \left( -A^2 f(r)^2 + A f(r) + 2 \right) b(r) + A^2 f(r)^2 r}{4 \kappa r^3}.
\end{equation}
Evaluating $p_t(r)$ at the throat $r_0$, we have
\begin{equation}
	p_t(r_0) = \frac{ r_0 \left( -A f(r_0) - 2 \right) b'(r_0) + \left( -A^2 f(r_0)^2 + A f(r_0) + 2 \right) r_0 + A^2 f(r_0)^2 r_0}{4 \kappa r_0^3}.
\end{equation}
Using the first EFE, the derivative of the shape function can be expressed as
\begin{equation}
	b'(r) = \kappa r^2 \rho(r) = \kappa r^2 \rho_S \frac{R \sin(\pi r / R)}{\pi r}, \label{EFE1}
\end{equation}
and, in particular at the throat,
\begin{equation}
	b'(r_0) = \kappa r_0 \rho_S \frac{R \sin(\pi r_0 / R)}{\pi} \simeq \kappa r_0^2 \rho_S < 1, \quad (R \gg r_0),
\end{equation}
where the last inequality ensures the flare-out condition.

To impose a vanishing transverse pressure at $r = R$, we expand $f(r)$ and $b(r)$ near the boundary:
\begin{equation}
	f(r) = f(R) + F(R) (r-R) + o((r-R)^2), \qquad F(R) = f'(r)|_{r=R},
\end{equation}
\begin{equation}
	b(r) = b(R) + b'(r)|_{r=R} (r-R) + o((r-R)^2) \simeq b(R) + o((r-R)^2),
\end{equation}
where we have used $\left( \ref{EFE1} \right)$ to neglect $b'(R)$ in the linear approximation. Then, the transverse pressure at the boundary reduces to
\begin{equation}
	p_t(R) = \frac{\left[ (R - b(R)) F(R) + f(R) \right] b(R)}{2 \kappa R^2 (R - b(R)) f(R)}.
\end{equation}
Setting $p_t(R) = 0$ gives
\begin{equation}
	F(R) = - \frac{f(R)}{R - b(R)},
\end{equation}
which ensures that both radial and transverse pressures vanish smoothly at the outer boundary.

A simple choice that satisfies this condition in the outer layer $r \in (R-\varepsilon, R]$ is
\begin{equation}
	f(r) = K \exp\left( - \frac{r}{R - b(R)} \right),
\end{equation}
where $K$ is a constant. Note that $f(r)$ cannot be uniquely determined outside this narrow boundary layer.  

In the following sections, we explore several alternative redshift profiles motivated by the polytropic $n=1$ energy density $\rho(r)$ in $\left( \ref{TF} \right)$, thereby enriching the set of potential traversable wormhole solutions.

\section{Cored-Inspired Redshift Function: Proposal I}\label{Sec5}

An alternative approach to defining the redshift function $\Phi(r)$ is inspired directly by the polytropic density profile itself. Instead of imposing an explicit EoS, we can consider a redshift function that mimics the cored behavior of the polytropic profile:
\begin{equation}
	\Phi(r) = \Phi_0 \frac{\sin(kr)}{kr} = \Phi_0 \frac{R \sin(\pi r / R)}{\pi r}, \label{Phi(r)TF}
\end{equation}
where $\Phi_0$ is a constant that will be fixed by the conditions at the throat $r_0$ and at the boundary $R$. By construction, this choice guarantees that $\Phi(R) = 0$, preventing the formation of an event horizon at the outer boundary.

Plugging the redshift function \eqref{Phi(r)TF} and the shape function \eqref{b(r)} into the second Einstein field equation (EFE) \eqref{pr0} yields the radial pressure
\begin{equation}
	p_r(r) = \frac{2\rho_S}{k^2 r^2} \left( \Phi_0 \cos^2(kr) + \frac{\sin^2(kr)}{k^2 r^2} \right) - \frac{\cos(kr)}{\kappa k^3 r^3} A(r) + \frac{\sin(kr) B(r) - kr C(r_0)}{k^4 r^4 \kappa},
\end{equation}
where
\begin{align}
	A(r) &= \kappa \rho_S \Phi_0 \left( 2 \sin(kr_0) - 2 \cos(kr_0) k r_0 - 4 \sin(kr) \right) + \kappa \rho_S kr + 2 \Phi_0 k^3 (r - r_0), \\
	B(r) &= \kappa \rho_S \left( 2 \Phi_0 \cos(kr_0) k r_0 - 2 \Phi_0 \sin(kr_0) - kr \right) - 2 \Phi_0 k^3 (r - r_0), 
\end{align}
and
\begin{align}
	C(r_0) = \kappa \rho_S \left( \cos(kr_0) k r_0 - \sin(kr_0) \right) + r_0 k^3.
\end{align}

To impose a vanishing radial pressure at the outer boundary, $p_r(R) = 0$, we substitute $k = \pi / R$ and solve for $\Phi_0$, obtaining
\begin{equation}
	\Phi_0 = \frac{\rho_S \kappa \left( \cos(\pi r_0 / R) R^2 \pi r_0 + \pi R^3 - \sin(\pi r_0 / R) R^3 \right) + r_0 \pi^3}{2 \rho_S \kappa \left( \cos(\pi r_0 / R) R^2 \pi r_0 + \pi R^3 - \sin(\pi r_0 / R) R^3 \right) - 2 \pi^3 R + 2 r_0 \pi^3}. \label{Phi0}
\end{equation}

Using the approximations
\begin{align}
	\cos(\pi r_0 / R) &\simeq 1 - \frac{1}{2} \left( \frac{\pi r_0}{R} \right)^2 + O\left( \left( \frac{\pi r_0}{R} \right)^4 \right), \\
	\sin(\pi r_0 / R) &\simeq \frac{\pi r_0}{R} - \frac{1}{6} \left( \frac{\pi r_0}{R} \right)^3 + O\left( \left( \frac{\pi r_0}{R} \right)^5 \right),
\end{align}
we can write a simplified form
\begin{equation}
	\Phi_0 \simeq -\frac{\pi^2 \kappa r_0^3 \rho_S - 3 R^3 \rho_S \kappa - 3 \pi^2 r_0}{2 \left( 3 R^3 \rho_S \kappa - \pi^2 \left( 3 (R - r_0) + \kappa r_0^3 \rho_S \right) \right)}.
\end{equation}

The third EFE gives the transverse pressure
\begin{align}
	p_t(r) &= \frac{\Phi_0^2 \rho_S}{r^2 k^2} \left( \cos^3(kr) - \frac{\sin^3(kr)}{r^3 k^3} \right) + A_1(r) \cos^2(kr) + \left( \frac{3 \Phi_0^2 \rho_S \sin^2(kr)}{k^4 r^4} + A_2(r) \right) \cos(kr) \nonumber \\
	& \quad + A_3(r) \frac{\sin^2(kr)}{r^5 k^5 \kappa} + A_4(r) \frac{\sin(kr)}{r^4 k^4 \kappa} + A_5(r),
\end{align}
where the functions $A_1(r), \dots, A_5(r)$ are defined in Appendix \ref{A5}. Evaluating at $r = R$, we obtain
\begin{equation}
	p_t(R) = - \frac{\Phi_0^2 \rho_S}{\pi^2} + A_1(R) - A_2(R) + A_5(R),
\end{equation}
with%
\begin{align}
	A_{1}\left(  R\right)    & =\left(  \frac{\Phi_{0}^{2}\left(  \sin\!\left(
		\pi r_{0}/R\right)  \pi R^{2}\kappa\rho_{S}-R\cos\!\left(  \pi r_{0}/R\right)
		\pi^{2}\kappa r_{0}\rho_{S}+\left(  \left(  R-r_{0}\right)  \pi^{3}\right)
		\right)  -3R^{3}\rho_{S}\kappa\pi\Phi_{0}}{R^{3}\pi^{3}\kappa}\right)  ,\\
	A_{2}\left(  R\right)    & =\frac{3\Phi_{0}\kappa\rho_{S}\left(  R^{2}%
		\cos\left(  \pi r_{0}/R\right)  \pi r_{0}-R^{3}\sin\left(  \pi r_{0}/R\right)
		\right)  -\left(  \left(  2R-3r_{0}\right)  \Phi_{0}\,\pi^{3}+R^{3}\pi\rho
		_{S}\kappa\right)  }{2\pi^{3}\kappa R^{3}},
\end{align}
and%
\begin{equation}
	A_{5}\left(  R\right)  =\frac{3\Phi_{0}\kappa\rho_{S}\pi R^{3}+\kappa\rho
		_{S}\left(  R^{2}\cos\left(  \pi r_{0}/R\right)  \pi r_{0}-R^{3}\sin\left(
		\pi r_{0}/R\right)  \right)  +r_{0}\pi^{3}}{2\pi^{3}\kappa R^{3}}.
\end{equation}

With the help of Eq.~$\left(  \ref{Phi0}\right)  $, $p_{t}\!\left(  R\right)  $
can be further reduced to obtain%
\begin{equation}
	p_{t}\!\left(  R\right)  =-\frac{\left(  \cos\!\left(  \pi r_{0}/R\right)
		R^{2}\pi r_{0}\rho_{S}\kappa+\pi R^{3}\kappa\rho_{S}-\sin\!\left(  \pi
		r_{0}/R\right)  R^{3}\rho_{S}\kappa+r_{0}\pi^{3}\right)  ^{2}}{2\pi^{3}%
		R^{3}\kappa\left(  \cos\!\left(  \pi r_{0}/R\right)  R^{2}\pi r_{0}\rho
		_{S}\kappa-\sin\!\left(  \pi r_{0}/R\right)  R^{3}\rho_{S}\kappa+\left(
		\left(  r_{0}-R\right)  \pi^{2}+R^{3}\kappa\rho_{S}\right)  \pi\right)  }%
\end{equation}
and, in the approximation described by Eqs.$\left(  \ref{AC}\right)  $ and
$\left(  \ref{AS}\right)  $, we can write%
\begin{equation}
	p_{t}\!\left(  R\right)  =-\frac{{\left(  \left(  3-\kappa r_{0}^{2}\rho
			_{S}\right)  r_{0}\pi^{2}+3R^{3}\kappa\rho_{S}\right)  }^{2}}{2\pi^{2}%
		R^{3}\left(  \left(  3-\kappa r_{0}^{2}\rho_{S}\right)  r_{0}\pi^{2}%
		+3R^{3}\kappa\rho_{S}\right)  \kappa}.
\end{equation}

To ensure $p_t(R) = 0$, the following condition on the boundary radius $R$ emerges:
\begin{equation}
	R = \left( \frac{\pi^2 r_0 (\kappa r_0^2 \rho_S - 3)}{3 \kappa \rho_S} \right)^{1/3}. \label{R}
\end{equation}
For a real solution, we must impose
\begin{equation}
	\rho_S > \frac{3}{\kappa r_0^2}.
\end{equation}

This inequality must be compatible with the upper bound derived in \eqref{UB}. Defining $x = R / r_0$, we find $x \in (1, 138)$. In particular, the allowed values satisfy
\begin{equation}
	\sqrt[3]{\frac{\pi^2}{3}} r_0 > R > r_0,
\end{equation}
which indicates that the outer boundary is very close to the throat. This motivates the consideration of alternative proposals for the redshift function to allow for a more extended wormhole geometry.

\section{Cored-Inspired Redshift Function: Proposal II}
\label{Sec6}

A natural generalization of Proposal I consists in introducing a constant
phase shift in the redshift function. This additional degree of freedom
allows for a richer class of cored profiles while preserving the regular
behavior at the throat and the vanishing of $\Phi(r)$ at the outer boundary.
We therefore consider
\begin{equation}
	\Phi(r)=\Phi_{0}\frac{\sin(kr+D)}{kr}
	=\Phi_{0}\frac{\sin(kr)\cos D+\sin D \cos(kr)}{kr},
	\label{Phi(r)II}
\end{equation}
where $D$ is a constant phase. This expression can be conveniently rewritten as
\begin{equation}
	\Phi(r)=\Phi_{1}\frac{\sin(kr)}{kr}
	+\Phi_{2}\frac{\cos(kr)}{kr},
\end{equation}
with $\Phi_{1}=\Phi_{0}\cos D$ and $\Phi_{2}=\Phi_{0}\sin D$. In this
representation, $\Phi_{1}$ and $\Phi_{2}$ are independent constants to be
fixed by the throat and boundary conditions. The previous proposal is
recovered in the limit $\Phi_{2}=0$.

Substituting Eq.~\eqref{Phi(r)II} into the second Einstein field equation
\eqref{pr0}, we obtain
\begin{align}
	p_{r}(r) &=
	\frac{2\rho_{S}\left(\Phi_{1}kr-2\Phi_{2}\right)\cos^{2}(kr)}{k^{3}r^{3}}
	+\frac{2\Phi_{1}\sin^{2}(kr)\rho_{S}}{k^{4}r^{4}}
	+B_{1r}(r)\frac{\cos(kr)}{\kappa k^{4}r^{4}}
	\nonumber\\
	&\quad
	+B_{2r}(r)\frac{\sin(kr)}{\kappa k^{4}r^{4}}
	+B_{3r}(r)\frac{\sin(kr)\cos(kr)}{k^{4}r^{4}}
	+\frac{B_{4r}(r)}{k^{3}r^{3}\kappa},
\end{align}
where the functions $B_{1r}(r),\dots,B_{4r}(r)$ are given in Appendix~\ref{A6}.

Evaluating at the external boundary $r=R$ (with $k=\pi/R$) yields
\begin{equation}
	p_{r}(R)=
	\frac{2\rho_{S}\left(\Phi_{1}kR-2\Phi_{2}\right)}{\pi^{3}}
	-\frac{B_{1}(R)}{\kappa\pi^{4}}
	+\frac{B_{4}(R)}{\pi^{3}\kappa}.
\end{equation}
Imposing $p_{r}(R)=0$ determines $\Phi_{1}$ as a function of $\Phi_{2}$,
\begin{equation}
	\Phi_{1}=\frac{\kappa R^{2}\rho_{S}\left(  2\Phi_{2}+\pi\right)  \left(
		r_{0}\pi\cos\!\left(  \pi r_{0}/R\right)  -R\sin\!\left(  \pi r_{0}/R\right)
		+R\right)  +\left(  r_{0}\pi^{3}-2\Phi_{2}\left(  R-r_{0}\right)  \pi
		^{2}\right)  \pi}{2\left(  \rho_{S}\kappa R^{2}\left(  \cos\!\left(  \pi
		r_{0}/R\right)  \pi r_{0}-\sin\!\left(  \pi r_{0}/R\right)  R+\right)
		+\left(  \left(  r_{0}-R\right)  \pi^{2}+R^{3}\rho_{S}\kappa\right)
		\pi\right)  \pi},
\end{equation}
which, under the approximations \eqref{AC} and \eqref{AS}, reduces to
\begin{equation}
	\Phi_{1}=
	\frac{\left(-\kappa r_{0}^{3}\rho_{S}+3r_{0}\right)\pi^{3}
		-6\Phi_{2}\left(\frac{1}{3}\kappa r_{0}^{3}\rho_{S}+R-r_{0}\right)\pi^{2}
		+3\pi R^{3}\rho_{S}\kappa
		+6\Phi_{2}\kappa\rho_{S}R^{3}}
	{\left(\left(6R^{3}\rho_{S}\kappa
		-2\kappa r_{0}^{3}\rho_{S}
		-3(R-r_{0})\right)\pi^{2}\right)\pi}.
	\label{Phi1II}
\end{equation}

\medskip
\noindent
\textbf{Transverse pressure.}
The third Einstein field equation leads to a lengthy but well-defined
expression,
\begin{eqnarray}
	p_{t}\!\left(  r\right)    & = &\rho_{S}\left(  \Phi_{1}^{2}k^{2}r^{2}-6\Phi_{1}%
	\Phi_{2}kr+3\Phi_{2}^{2}\right)  \frac{\cos^{3}\!\left(  kr\right)  }%
	{k^{4}r^{4}}-\Phi_{1}^{2}\rho_{S}\frac{\sin^{3}\!\left(  kr\right)  }%
	{k^{5}r^{5}}+B_{1t}\left(  r\right)  \frac{\sin^{2}\!\left(  kr\right)
	}{2\kappa k^{5}r^{5}}+B_{2t}\left(  r\right)  \frac{\cos^{2}\!\left(
		kr\right)  }{2\kappa k^{5}r^{5}}\nonumber\\
	&& +B_{3t}\left(  r\right)  \frac{\cos\!\left(  kr\right)  \sin^{2}\!\left(
		kr\right)  }{k^{5}r^{5}}+B_{4t}\left(  r\right)  \frac{\sin\!\left(
		kr\right)  \cos^{2}\!\left(  kr\right)  }{k^{5}r^{5}}+B_{5t}\left(  r\right)
	\frac{\sin\!\left(  kr\right)  \cos\!\left(  kr\right)  }{2\kappa k^{5}r^{5}%
	}+B_{6t}\left(  r\right)  \frac{\cos\!\left(  kr\right)  }{2r^{4}k^{4}\kappa
	}\nonumber\\
	&& +B_{7t}\left(  r\right)  \frac{\sin\!\left(  kr\right)  }{2r^{4}k^{4}\kappa
	}+\frac{B_{8t}\left(  r\right)  }{2k^{4}r^{4}\kappa},
\end{eqnarray}
where the explicit forms of $B_{it}(r)$ are given in Appendix~\ref{A6}.  

Using again the approximations \eqref{AC} and \eqref{AS}, the boundary value
$r=R$ becomes
\begin{eqnarray}
	p_{t}\!\left(  R\right) & = &\frac{\left(  \left(  2\kappa r_{0}^{3}%
		\rho+6R-6r_{0}\right)  \pi^{4}-6\pi^{2}R^{3}\rho\kappa\right)  \Phi_{1}^{2}%
	}{6R^{3}\pi^{4}\kappa}+\frac{\left(  \left(  2\kappa r_{0}^{3}\rho
		+6R-6r_{0}\right)  \pi^{2}-6R^{3}\rho\kappa\right)  \Phi_{2}^{2}}{6R^{3}%
		\pi^{4}\kappa}
	\nonumber\\
	&&+\frac{\left(  \left(  3\kappa r_{0}^{3}\rho+6R-9r_{0}\right)  \pi^{4}-\left(
		4\kappa r_{0}^{3}\rho+12\left(  R-r_{0}\right)  \right)  \Phi_{2}\pi^{3}%
		-9\pi^{2}R^{3}\rho\kappa+12\Phi_{2}\pi R^{3}\kappa\rho\right)  \Phi_{1}%
	}{6R^{3}\pi^{4}\kappa}
	\nonumber\\
	&&
	+\frac{\left(  \left(  2\kappa r_{0}^{3}\rho+6\left(  R-r_{0}\right)  \right)
		\pi^{5}-3\left(  \rho\left(  2R^{3}+r_{0}^{3}\right)  \kappa+2R-3r_{0}\right)
		\pi^{3}+9\pi R^{3}\rho\kappa\right)  \Phi_{2}}{6\pi^{4}R^{3}\kappa}
	\nonumber\\
	&&+\frac{\left(  -\kappa r_{0}^{3}\rho+3r_{0}\right)  \pi^{4}+3\pi^{2}R^{3}%
		\rho\kappa}{6R^{3}\pi^{4}\kappa},
	\label{pt(R)II}%
\end{eqnarray}
which depends quadratically on $\Phi_{1}$ and $\Phi_{2}$.

Imposing $p_{t}(R)=0$ and substituting Eq.~\eqref{Phi1II}, we obtain
\begin{equation}
	\Phi_{2}
	=-\frac{\left(\left(3r_{0}-\kappa r_{0}^{3}\rho_{S}\right)\pi^{2}
		+3R^{3}\rho_{S}\kappa\right)^{2}}
	{2\pi\left(\left(3R^{3}\rho_{S}\kappa
		-\kappa r_{0}^{3}\rho_{S}
		-3(R-r_{0})\right)\pi^{2}\right)^{2}}.
	\label{Phi2II}
\end{equation}
Once $\Phi_{2}$ is known, $\Phi_{1}$ follows directly from
Eq.~\eqref{Phi1II}.

Note that $\Phi_{2}$ vanishes when we use the value determined by Eq.~$\left(
\ref{R}\right)  $ into Eq.~(  \ref{Phi2II}). Now, we can use the
value of $\Phi_{2}$ of Eq.~(  \ref{Phi2II}) to compute the value
of $\Phi_{1}$, which is given by
\begin{eqnarray}
	\Phi_{1}  & = &\frac{\left(  \left(  3r_{0}-\kappa r_{0}^{3}\rho_{S}\right)
		\pi^{2}+3R^{3}\rho_{S}\kappa\right)  \left(  3r_{0}-\kappa r_{0}^{3}\rho
		_{S}-3R\right)  \pi^{4}}{6{\left(  3R^{3}\rho_{S}\kappa-\left(  \kappa
			r_{0}^{3}\rho_{S}+3\left(  R-r_{0}\right)  \right)  \pi^{2}\right)  }^{2}%
		\pi^{2}}\nonumber\\
	&& +\frac{3\left(  \rho_{S}\left(  3R^{3}+r_{0}^{3}\right)  \kappa
		-3r_{0}\right)  \pi^{2}-9R^{3}\rho_{S}\kappa}{6\pi^{2}{\left(  3R^{3}\rho
			_{S}\kappa-\left(  \kappa r_{0}^{3}\rho_{S}+3\left(  R-r_{0}\right)  \right)
			\pi^{2}\right)  }^{2}}.
\end{eqnarray}

The validity domain of $\Phi_{1}$ and $\Phi_{2}$ is satisfied if
\begin{equation}
	R\neq\frac{\sqrt[3]{12\pi^{2}}\left(  \sqrt[3]{12\pi^{2}}+\left(  \kappa
		\rho_{S}\right)  ^{\frac{1}{3}}\left(  3\left(  \kappa\rho_{S}\right)
		^{\frac{3}{2}}r_{0}^{3}+\sqrt{9\kappa^{3}r_{0}^{6}\rho_{S}^{3}-54\kappa
			^{2}r_{0}^{4}\rho_{S}^{2}+81r_{0}^{2}\rho_{S}\kappa-12\pi^{2}}-9\sqrt
		{\kappa\rho_{S}}r_{0}\right)  ^{\frac{2}{3}}\right)  }{6\left(  \kappa\rho
		_{S}\right)  ^{\frac{5}{6}}\left(  3\left(  \kappa\rho_{S}\right)  ^{\frac
			{3}{2}}r_{0}^{3}+\sqrt{9\kappa^{3}r_{0}^{6}\rho_{S}^{3}-54\kappa^{2}r_{0}%
			^{4}\rho_{S}^{2}+81r_{0}^{2}\rho_{S}\kappa-12\pi^{2}}-9\sqrt{\kappa\rho_{S}%
		}r_{0}\right)  ^{\frac{1}{3}}}\,.
\end{equation}

A further simplification arises by fixing
\begin{equation}
	\rho_{S}=\frac{3}{\kappa r_{0}^{2}},
	\label{rhoSII}
\end{equation}
for which
\begin{equation}
	\Phi_{2}=-\frac{9R^{4}}{2(3R^{2}-r_{0}^{2}\pi^{2})^{2}\pi},
	\qquad
	\Phi_{1}=-\frac{3R^{2}\left(\pi^{4}r_{0}^{2}
		+3R^{2}(1-\pi^{2})\right)}
	{2(3R^{2}-\pi^{2}r_{0}^{2})^{2}\pi^{2}}.
\end{equation}
In this case, consistency requires
\begin{equation}
	R>r_{0}\frac{\sqrt{3}\pi}{3}.
\end{equation}

\medskip
\noindent
The freedom in fixing $\rho_{S}$ through Eq.~\eqref{rhoSII} is allowed
because both $p_{r}(R)$ and $p_{t}(R)$ vanish, leaving residual parameter
freedom that can be exploited to simplify the redshift coefficients.
Compared to Proposal I, the present construction provides a broader class of
cored-inspired redshift functions, at the price of additional algebraic
complexity. 

In the next section, we shall adopt a related strategy by imposing a
cored-inspired profile directly on $\Phi'(r)$ rather than on $\Phi(r)$,
which will allow for further analytical control of the boundary conditions.

\section{Cored Inspired Redshift Function: Proposal III}\label{Sec7}

This third proposal differs conceptually from the constructions
introduced in Sections \ref{Sec5} and \ref{Sec6}. Instead of prescribing
directly the functional form of $\Phi(r)$, we now assume that its radial
derivative inherits the cored structure of the polytropic $n=1$ profile.
In this way, the regular behavior is imposed at the level of the
gravitational redshift gradient, while $\Phi(r)$ itself follows from
integration. Concretely, we assume
\begin{equation}
	\Phi^{\prime}\left(r\right)
	=\Phi_{3}\frac{\sin\left(kr\right)}{kr}
	+\Phi_{4}\frac{\cos\left(kr\right)}{kr},
	\label{Phi(r)III}
\end{equation}
where $\Phi_{3}$ and $\Phi_{4}$ are constants with dimensions
$\left[L^{-1}\right]$, to be fixed by the throat radius $r_{0}$
and by the boundary $R$.

Substituting Eq.~\eqref{Phi(r)III} into the second Einstein field
equation \eqref{pr0}, one finds
\begin{eqnarray}
	p_{r}\!\left(r\right)
	&=&\frac{2\rho_{S}\Phi_{4}\cos^{2}\!\left(kr\right)}{k^{3}r^{2}}
	-\frac{2\rho_{S}\Phi_{3}\sin^{2}\!\left(kr\right)}{k^{4}r^{3}}
	+\frac{2\rho_{S}\left(kr\Phi_{3}-\Phi_{4}\right)
		\sin\!\left(kr\right)\cos\!\left(kr\right)}{k^{4}r^{3}}
	\nonumber\\
	&&
	+C_{1r}\!\left(r\right)\frac{\cos\!\left(kr\right)}{k^{4}r^{3}\kappa}
	+C_{2r}\!\left(r\right)\frac{\sin\!\left(kr\right)}{k^{4}r^{3}\kappa}
	-\frac{\cos\!\left(kr_{0}\right)k\kappa r_{0}\rho_{S}
		+r_{0}k^{3}
		-\sin\!\left(kr_{0}\right)\kappa\rho_{S}}
	{k^{3}r^{3}\kappa},
\end{eqnarray}
where
\begin{equation}
	C_{1r}\!\left(r\right)
	=2\kappa\rho_{S}\Phi_{4}
	\left(\sin\!\left(kr_{0}\right)
	-\cos\!\left(kr_{0}\right)kr_{0}\right)
	+2\left(2\Phi_{4}(r-r_{0})k+r\rho_{S}\kappa\right)k^{2},
\end{equation}
and
\begin{equation}
	C_{2r}\!\left(r\right)
	=-2\cos\!\left(kr_{0}\right)k\kappa r_{0}\rho_{S}\Phi_{3}
	+2\sin\!\left(kr_{0}\right)\kappa\rho_{S}\Phi_{3}
	+\left(2\Phi_{3}(r-r_{0})k^{2}-\kappa\rho_{S}\right)k.
\end{equation}

As in the previous proposals, we impose the physically motivated
condition that the radial pressure vanishes at the boundary.
Setting $k=\pi/R$, we obtain
\begin{eqnarray}
	p_{r}\!\left(R\right)
	&=&\frac{\pi\left(2R^{3}\kappa r_{0}\rho_{S}\Phi_{4}
		-\pi R^{2}\kappa r_{0}\rho_{S}\right)
		\cos\!\left(\pi r_{0}/R\right)}{\pi^{4}R^{3}\kappa}
	-\frac{R\left(2R^{3}\rho_{S}\kappa\Phi_{4}
		-\pi R^{2}\kappa\rho_{S}\right)
		\sin\!\left(\pi r_{0}/R\right)}{\pi^{4}R^{3}\kappa}
	\nonumber\\
	&&
	+\frac{2\pi\kappa R^{4}\rho_{S}\Phi_{4}
		-\pi^{2}\left(\rho_{S}\kappa R^{2}
		+2\Phi_{4}(R-r_{0})\pi\right)R
		-\pi^{4}r_{0}}
	{\pi^{4}R^{3}\kappa}.
\end{eqnarray}

Using the approximations \eqref{AC} and \eqref{AS},
the condition $p_{r}(R)=0$ yields
\begin{equation}
	\Phi_{4}
	=\frac{3\pi R^{3}\kappa\rho_{S}
		-\left(\kappa r_{0}^{3}\rho_{S}-3r_{0}\right)\pi^{3}}
	{2R\pi^{2}\left(3R^{3}\rho_{S}\kappa
		-\kappa r_{0}^{3}\rho_{S}
		+3\left(R-r_{0}\right)\right)}.
	\label{Phi4}
\end{equation}

The third Einstein field equation \eqref{pt0} leads to
\begin{eqnarray}
	p_{t}\!\left(  r\right)  &=&\frac{\rho_{S}\Phi_{4}^{2}\cos^{3}\!\left(
		kr\right)  }{k^{4}r^{2}}-\frac{\rho_{S}\Phi_{3}^{2}\sin^{3}\!\left(
		kr\right)  }{k^{5}r^{3}}
	\nonumber\\
	&&+\frac{\rho_{S}\Phi_{3}\left(  kr\Phi_{3}-2\Phi_{4}\right)  \cos\!\left(
		kr\right)  }{k^{5}r^{3}}\sin^{2}\!\left(  kr\right)  +\frac{\Phi_{4}\rho
		_{S}\left(  2kr\Phi_{3}-\Phi_{4}\right)  \sin\!\left(  kr\right)  }{k^{5}%
		r^{3}}\cos\!\left(  kr\right)  ^{2}
	\nonumber\\
	&&+C_{1t}\left(  r\right)  \frac{\sin^{2}\!\left(  kr\right)  }{2k^{5}%
		r^{3}\kappa}+C_{2t}\left(  r\right)  \frac{\cos\!\left(  kr\right)  ^{2}%
	}{2k^{5}r^{3}\kappa}+C_{3t}\left(  r\right)  \frac{\sin\!\left(  kr\right)
	}{2k^{5}r^{3}\kappa}+C_{4t}\left(  r\right)  \frac{\sin\!\left(  kr\right)
	}{2k^{4}r^{3}\kappa}+C_{5t}\left(  r\right)  \frac{\cos\!\left(  kr\right)
	}{2k^{4}r^{3}\kappa}
	\nonumber\\
	&&+\frac{-k\kappa r^{2}\rho_{S}\Phi_{3}+\cos\!\left(  kr_{0}\right)  k\kappa
		r_{0}\rho_{S}+r_{0}k^{3}+2r\rho_{S}\kappa\Phi_{4}-\sin\!\left(  kr_{0}\right)
		\kappa\rho_{S}}{2k^{3}r^{3}\kappa},
\end{eqnarray}
where the functions $C_{it}(r)$ are defined in
Appendix~\ref{A7}.

After substituting $k=\pi/R$ and employing again
Eqs.~\eqref{AC} and \eqref{AS}, together with Eq.~\eqref{Phi4},
the condition $p_{t}(R)=0$ determines $\Phi_{3}$,
\begin{equation}
	\Phi_{3}
	=\frac{\pi^{2}\left(3R^{3}\kappa\rho_{S}
		+\pi^{2}\left(3r_{0}-\kappa r_{0}^{3}\rho_{S}\right)\right)}
	{2\left(\left(3R^{3}\kappa\rho_{S}
		-\kappa r_{0}^{3}\rho_{S}
		-3\left(R-r_{0}\right)\right)\pi^{2}\right)^{2}}.
\end{equation}
The coefficients $\Phi_{3}$ and $\Phi_{4}$ share the same
domain of validity found for $\Phi_{1}$ and $\Phi_{2}$,
ensuring consistency among the three proposals.

A final remark on the choice \eqref{Phi(r)III} is in order.
Solving the differential equation
\begin{equation}
	\Phi^{\prime}\left(r\right)
	=\Phi_{3}\frac{\sin\left(kr\right)}{kr}
	+\Phi_{4}\frac{\cos\left(kr\right)}{kr},
\end{equation}
one obtains
\begin{equation}
	\Phi\left(r\right)
	=\Phi_{3}\frac{\operatorname{Si}\left(kr\right)}{k}
	+\Phi_{4}\frac{\operatorname{Ci}\left(kr\right)}{k}
	+K,
\end{equation}
where $K$ is an integration constant,
\begin{equation}
	\operatorname{Si}(x)
	=\int_{0}^{x}\frac{\sin t}{t}\,dt
\end{equation}
is the sine integral, and
\begin{equation}
	\operatorname{Ci}(x)
	=\int_{0}^{x}\frac{\cos t}{t}\,dt
\end{equation}
is the cosine integral.

Therefore, in this proposal the redshift function is expressed
in terms of special functions, leading to a smoother and more
flexible profile. This construction preserves the cored behavior
at small $r$, while providing additional analytical freedom
through the integration constant $K$, which can be fixed
by requiring $\Phi(R)=0$.

\section{Cored Inspired Redshift Function: Proposal IV}\label{Sec8}

A further generalization of the previous construction can be obtained by introducing an explicit reference to the throat scale directly in the definition of $\Phi'(r)$. In contrast with Proposal III, where both oscillatory contributions were centered at the origin, we now subtract a constant term proportional to the throat value. This choice allows us to control the behavior of the redshift gradient at $r=r_{0}$, effectively introducing an ``initial phase'' anchored at the throat.

We therefore consider
\begin{equation}
	\Phi^{\prime}\left(r\right)
	=\Phi_{5}\frac{\sin\left(kr\right)}{kr}
	-\Phi_{6}\frac{\sin\left(kr_{0}\right)}{kr_{0}},
	\label{Phi(r)IV}
\end{equation}
where $\Phi_{5}$ and $\Phi_{6}$ are constants with dimensions
$\left[L^{-1}\right]$, to be determined by the throat radius $r_{0}$
and by the boundary $R$. The subtraction term ensures that the
redshift gradient can be tuned to vanish or take a prescribed value
at $r=r_{0}$, providing additional flexibility with respect to the
previous proposals.

Substituting Eq.~\eqref{Phi(r)IV} into the second Einstein field
equation \eqref{pr0}, we obtain
\begin{eqnarray}
	p_{r}\!\left(r\right)
	&=&\frac{2\Phi_{5}\rho_{S}\cos\!\left(kr\right)\sin\!\left(kr\right)}
	{k^{3}r^{2}}
	-\frac{2\rho_{S}\Phi_{5}\sin^{2}\!\left(kr\right)}{r^{3}k^{4}}
	\nonumber\\
	&&
	+kr\rho_{S}\kappa\left(kr_{0}-2r\Phi_{6}
	\sin\!\left(kr_{0}\right)\right)
	\frac{\cos\!\left(kr\right)}{\kappa k^{4}r^{3}r_{0}}
	\nonumber\\
	&&
	+D_{1r}\!\left(r\right)
	\frac{\sin\!\left(kr\right)}{\kappa k^{4}r^{3}r_{0}}
	+\frac{\kappa\rho_{S}kr_{0}
		\left(\sin\!\left(kr_{0}\right)
		-\cos\!\left(kr_{0}\right)kr_{0}\right)
		-k^{4}r_{0}^{2}}
	{r^{3}k^{4}r_{0}\kappa}
	\nonumber\\
	&&
	+\frac{2\rho_{S}\kappa\sin\!\left(kr_{0}\right)r
		\left(\cos\!\left(kr_{0}\right)kr_{0}
		-\sin\!\left(kr_{0}\right)
		-k^{3}r(r-r_{0})
		\sin\!\left(kr_{0}\right)\right)\Phi_{6}}
	{r^{3}k^{4}r_{0}\kappa},
\end{eqnarray}
where
\begin{equation}
	D_{1r}\!\left(r\right)
	=\Big(
	2\rho_{S}\kappa\sin\!\left(kr_{0}\right)
	\left(r\Phi_{6}+r_{0}\Phi_{5}\right)
	+4k^{3}r_{0}\Phi_{5}(r-r_{0})
	-2\cos\!\left(kr_{0}\right)r_{0}\rho_{S}\kappa\Phi_{5}
	-k\kappa\rho_{S}
	\Big).
\end{equation}

As before, we require that the radial pressure vanishes at the
boundary $r=R$. Setting $k=\pi/R$ and using the approximations
\eqref{AC} and \eqref{AS}, we obtain
\begin{eqnarray}
	p_{r}\!\left(R\right)
	&=&-2\sin^{2}\!\left(\pi r_{0}/R\right)
	\kappa R^{5}\rho_{S}\Phi_{6}
	\nonumber\\
	&&
	-\pi^{2}\left(
	\cos\!\left(\pi r_{0}/R\right)R^{2}r_{0}\rho_{S}\kappa
	+R^{3}\rho_{S}\kappa
	+\pi^{2}r_{0}
	\right)r_{0}
	\nonumber\\
	&&
	+2R\pi\Big(
	R^{3}\cos\!\left(\pi r_{0}/R\right)r_{0}\rho_{S}\kappa\Phi_{6}
	+R^{4}\rho_{S}\kappa\Phi_{6}
	+\tfrac{1}{2}r_{0}\rho_{S}\kappa R^{2}
	-R\pi^{2}\Phi_{6}(R-r_{0})
	\Big)
	\sin\!\left(\pi r_{0}/R\right).
\end{eqnarray}

Solving the condition $p_{r}(R)=0$ with respect to $\Phi_{6}$,
one finds
\begin{equation}
	\Phi_{6}
	=\frac{3\left(\left(3r_{0}-\kappa r_{0}^{3}\rho_{S}\right)\pi^{2}
		+3R^{3}\rho_{S}\kappa\right)R}
	{\left(6R^{2}-\pi^{2}r_{0}^{2}\right)
		\left(\left(3R^{3}\rho_{S}\kappa
		-\kappa r_{0}^{3}\rho_{S}
		-3(R-r_{0})\right)\pi^{2}\right)}.
	\label{Phi6}
\end{equation}

The third Einstein field equation \eqref{pt0} leads to a lengthy
expression for $p_{t}(r)$,
\begin{eqnarray}
	p_{t}\!\left(  r\right) &=&\frac{3\rho_{S}\Phi_{5}\cos^{2}\!\left(  kr\right)
	}{2k^{2}r}-\frac{\rho_{S}\Phi_{5}^{2}\sin\!^{3}\left(  kr\right)  }{r^{3}%
		k^{5}}+\frac{\rho_{S}\Phi_{5}^{2}\sin^{2}\!\left(  kr\right)  \cos\!\left(
		kr\right)  }{r^{2}k^{4}}
	+D_{1t}\left(  r\right)  \frac{\sin^{2}\!\left(  kr\right)  }{2r^{3}k^{5}%
		r_{0}^{2}\kappa}
	\nonumber\\
	&&
	-\frac{\sin\!\left(  kr\right)  \cos\!\left(  kr\right)
	}{2r^{3}k^{5}r_{0}^{2}}\Phi_{5}\rho_{S}krr_{0}\left(  4r\Phi_{6}\sin\!\left(
	kr_{0}\right)  +3kr_{0}\right)  
	+\frac{\sin\!\left(  kr\right)  }{k^{5}r_{0}^{2}r}\left(  \frac{D_{2t}\left(
		r\right)  }{2r^{2}\kappa}+\frac{D_{3t}\left(  r\right)  }{2\kappa k^{5}%
		r^{3}r_{0}^{2}}-2\Phi_{6}^{2}\rho_{S}\sin^{2}\!\left(  kr_{0}\right)  \right)
	\nonumber\\
	&&+\frac{D_{4t}\left(  r\right)  \cos\!\left(  kr\right)  }{2k^{5}r_{0}^{2}%
		r^{3}\kappa}+\frac{D_{5t}\left(  r\right)  \cos\!\left(  kr\right)  }%
	{k^{5}r_{0}^{2}r^{3}\kappa}+\frac{D_{6t}\left(  r\right)  \cos\!\left(
		kr\right)  }{2k^{5}r_{0}^{2}r^{3}\kappa}
	-\rho_{S}\frac{\Phi_{5}}{2rk^{2}}
	\nonumber\\
	&&
	+\frac{r_{0}^{2}\left(  \cos\left(
		kr_{0}\right)  r_{0}\rho_{S}\kappa+k^{2}r_{0}\right)  k^{3}-\sin\left(
		kr_{0}\right)  k^{2}\kappa r_{0}^{2}\rho_{S}}{2k^{5}r_{0}^{2}r^{3}\kappa}%
\end{eqnarray}
where the functions $D_{it}(r)$ are defined in
Appendix~\ref{A8}.

Evaluating at $r=R$ (with $k=\pi/R$), one finds that
$p_{t}(R)$ is quadratic in $\Phi_{6}$ and linear in $\Phi_{5}$.
Imposing the boundary condition $p_{t}(R)=0$ and using
Eq.~\eqref{Phi6}, we can solve for $\Phi_{5}$,
\begin{equation}
	\Phi_{5}
	=-\frac{\left(
		3R^{3}\rho_{S}\kappa
		-\left(\kappa r_{0}^{3}\rho_{S}+6R-3r_{0}\right)\pi^{2}
		\right)
		\left(
		3R^{3}\rho_{S}\kappa
		-\left(\kappa r_{0}^{3}\rho_{S}-3r_{0}\right)\pi^{2}
		\right)}
	{6R\left(
		\left(
		3R^{3}\rho_{S}\kappa
		-\kappa r_{0}^{3}\rho_{S}
		-3(R-r_{0})
		\right)\pi^{2}
		\right)^{2}}.
\end{equation}

In summary, Proposal IV introduces an additional throat-dependent phase in the redshift gradient, allowing for a refined control of the geometry near $r_{0}$. As in the previous constructions, the full set of boundary conditions $p_{r}(R)=0$ and $p_{t}(R)=0$ uniquely fixes the coefficients $\Phi_{5}$ and $\Phi_{6}$. Compared to Proposals III and II, the present model provides greater flexibility in shaping the intermediate region between the throat and the boundary, while preserving the cored structure characteristic of the polytropic profile.

\section{Redshift Function Induced by the Third EFE}\label{Sec9}

In this section we adopt a different perspective with respect to the previous constructions. Instead of prescribing {\it a priori} a functional form for $\Phi(r)$ or $\Phi'(r)$, we extract suitable profiles directly from the third Einstein field equation~\eqref{pt0}. 
This strategy is not restricted to the present polytropic $n=1$ shape function, but can in principle be applied to any wormhole geometry once the shape function $b(r)$ is specified. The idea is to identify substructures within Eq.~\eqref{pt0} that can be isolated and solved as differential equations for $\Phi(r)$, thereby generating self-consistent redshift functions.

\subsection{First Proposal}

A first possibility arises by observing that Eq.~\eqref{pt0} contains,
as a homogeneous sector, the nonlinear differential equation
\begin{equation}
	\Phi''(r)+\left(\Phi'(r)\right)^{2}=0.
\end{equation}
This equation can be solved independently of the specific form of
$b(r)$ and admits the logarithmic solution
\begin{equation}
	\Phi(r)=\ln\!\left(c_{1}r+c_{2}\right),
	\label{Phi(r)A}
\end{equation}
where $c_{1}$ and $c_{2}$ are integration constants.
The logarithmic profile guarantees regularity provided
$c_{1}r+c_{2}>0$ in the interval $[r_{0},R]$.

Substituting Eq.~\eqref{Phi(r)A} back into the full third EFE,
the transverse pressure can be written as
\begin{equation}
	\kappa p_{t}(r)=
	\Bigg\{
	\left(1-\frac{b(r)}{r}\right)\frac{\Phi'(r)}{r}
	-\frac{b'(r)r-b(r)}{2r^{2}}
	\left(\Phi'(r)+\frac{1}{r}\right)
	\Bigg\}.
	\label{ptA}
\end{equation}

Recalling that the shape function is given by Eq.~\eqref{b(r)}
and setting $k=\pi/R$, Eq.~\eqref{ptA} becomes
\begin{eqnarray}
	p_{t}(r)
	&=&\frac{
		R\left(\left(R^{2}-\pi^{2}r^{2}\right)c_{2}
		-2r^{3}\pi^{2}c_{1}\right)
		\kappa\rho_{S}\sin\!\left(\pi r/R\right)
		+\cos\!\left(\pi r_{0}/R\right)\pi R^{2}c_{2}
		\kappa r_{0}\rho_{S}
	}
	{2\left(c_{1}r+c_{2}\right)\pi^{3}r^{3}\kappa}
	\nonumber\\
	&&
	-\frac{
		\sin\!\left(\pi r_{0}/R\right)R^{3}c_{2}\kappa\rho_{S}
		+\pi\left(
		\cos\!\left(\pi r/R\right)R^{2}c_{2}\kappa r\rho_{S}
		-2r^{2}c_{1}\pi^{2}
		-c_{2}\pi^{2}r_{0}
		\right)
	}
	{2\left(c_{1}r+c_{2}\right)\pi^{3}r^{3}\kappa}.
\end{eqnarray}

Using the approximations \eqref{AC} and \eqref{AS},
the boundary value simplifies considerably and reads
\begin{equation}
	p_{t}(R)
	=\frac{\left(3c_{1}R+2c_{2}\right)c_{1}}
	{R\kappa\left(c_{1}R+c_{2}\right)\left(3c_{1}R+c_{2}\right)}.
\end{equation}
The condition $p_{t}(R)=0$ can therefore be satisfied by choosing
\begin{equation}
	c_{2}=-\frac{3}{2}c_{1}R.
	\label{pt(R)A}
\end{equation}

With this relation, the radial pressure becomes
\begin{gather}
	p_{r}(r)
	=
	R^{2}\frac{(3c_{1}r+c_{2})}
	{(c_{1}r+c_{2})\pi^{3}r^{3}\kappa}
	\left(
	R\kappa\rho_{S}\sin\!\left(\pi r_{0}/R\right)
	+\pi\kappa r\rho_{S}\cos\!\left(\pi r/R\right)
	-\pi\kappa\rho_{S}r_{0}
	\cos\!\left(\pi r_{0}/R\right)
	\right)
	\nonumber\\
	+\frac{
		2\pi^{3}\left(
		c_{1}r^{2}
		-\frac{3}{2}c_{1}rr_{0}
		-\frac{1}{2}c_{2}r_{0}
		-R^{3}(3c_{1}r+c_{2})
		\kappa\rho_{S}\sin\!\left(\pi r/R\right)
		\right)
	}
	{(c_{1}r+c_{2})\pi^{3}r^{3}\kappa}.
\end{gather}

Evaluating again at $r=R$ and employing
Eqs.~\eqref{AC}, \eqref{AS}, together with
Eq.~\eqref{pt(R)A}, we find
\begin{equation}
	p_{r}(R)
	=\frac{\left(3r_{0}-\kappa r_{0}^{3}\rho_{S}-4R\right)\pi^{2}
		+3\rho_{S}R^{3}\kappa}
	{R^{3}\pi^{2}\kappa}.
\end{equation}

The boundary condition $p_{r}(R)=0$ then fixes the
polytropic density parameter as
\begin{equation}
	\rho_{S}
	=\frac{4\pi^{2}R-3\pi^{2}r_{0}}
	{\kappa\left(3R^{3}-\pi^{2}r_{0}^{3}\right)}.
\end{equation}
Positivity of $\rho_{S}$ requires
\begin{equation}
	R>\sqrt[3]{\frac{\pi^{2}\kappa}{3}}\,r_{0},
\end{equation}
and compatibility with the previously derived upper bound
\eqref{UB} further implies $R>3.06\,r_{0}$.

Thus, this first proposal illustrates how a viable
redshift function can be generated directly from the structure
of the third EFE. The logarithmic profile emerges naturally
from the homogeneous sector of the equation and, once the
boundary conditions are imposed, uniquely determines the
matter parameter $\rho_{S}$. This approach highlights the
flexibility of the formalism and opens the way to additional
redshift constructions derived from different sectors of
Eq.~\eqref{pt0}.

\subsection{Second Proposal}

An alternative possibility can be obtained by isolating a different sector of Eq.~\eqref{pt0}. In particular, we may impose the first-order differential equation
\begin{equation}
	\Phi'(r)+\frac{1}{r}=0,
\end{equation}
whose general solution is
\begin{equation}
	\Phi(r)=-\ln(r)+c_{3},
	\label{Phi(r)B}
\end{equation}
with $c_{3}$ an integration constant. 
This choice corresponds to a redshift function that decreases
logarithmically with the radial coordinate and remains regular
in the interval $(r_{0},R]$.

Substituting Eq.~\eqref{Phi(r)B} into the remaining part of the
third EFE \eqref{pt0}, the transverse pressure simplifies
considerably and becomes
\begin{equation}
	p_{t}(r)=\frac{r-b(r)}{\kappa r^{3}}.
\end{equation}
Since $b(r_{0})=r_{0}$, it immediately follows that
$p_{t}(r_{0})=0$, namely the transverse pressure vanishes
at the throat.

Evaluating the same expression at the boundary $r=R$, and using
the approximations \eqref{AC} and \eqref{AS}, we obtain
\begin{equation}
	p_{t}(R)=
	\frac{
		3\pi^{2}(R-r_{0})
		-\kappa\rho_{S}\left(3R^{3}-\pi^{2}r_{0}^{3}\right)
	}
	{3\kappa\pi^{2}R^{3}}.
\end{equation}
The condition $p_{t}(R)=0$ fixes the density parameter as
\begin{equation}
	\rho_{S}
	=\frac{3\pi^{2}(R-r_{0})}
	{\kappa\left(3R^{3}-\pi^{2}r_{0}^{3}\right)}.
	\label{rhoB}
\end{equation}
Positivity of $\rho_{S}$ once again requires
\begin{equation}
	R>\sqrt[3]{\frac{\pi^{2}\kappa}{3}}\,r_{0}.
\end{equation}

For the radial pressure evaluated at $r=R$ we find
\begin{equation}
	p_{r}(R)=
	\frac{
		3\rho_{S}\kappa R^{3}
		+\pi^{2}\left(3r_{0}-6R-\kappa\rho_{S}r_{0}^{3}\right)
	}
	{3\pi^{2}R^{3}\kappa}.
\end{equation}
Substituting Eq.~\eqref{rhoB} into the above expression,
the result simplifies drastically to
\begin{equation}
	p_{r}(R)=-\frac{1}{\kappa R^{2}},
\end{equation}
which is strictly negative and, in particular, never vanishes.

Therefore, although this proposal yields a simple and
analytically tractable redshift function, it fails to satisfy
the physically required boundary condition $p_{r}(R)=0$.
For this reason, the present case must be discarded as a
viable configuration.

\subsection{Third Proposal}

A further differential structure that can be isolated from the
third EFE \eqref{pt0} is
\begin{equation}
	\Phi''(r)+\frac{1}{r}\Phi'(r)=0,
\end{equation}
whose general solution is
\begin{equation}
	\Phi(r)=c_{5}\ln(r)+c_{4},
	\label{Phi(r)C}
\end{equation}
with $c_{4}$ and $c_{5}$ integration constants.
This profile generalizes the previous logarithmic ansatz and
reduces to a constant redshift function when $c_{5}=0$.

Substituting Eq.~\eqref{Phi(r)C} into Eq.~\eqref{pt0},
the transverse pressure takes the form
\begin{equation}
	p_{t}(r)=
	\frac{1}{2r^{3}\kappa}
	\left[
	2c_{5}^{2}r
	-r(c_{5}+1)b'(r)
	+b(r)\left(1+c_{5}-2c_{5}^{2}\right)
	\right].
\end{equation}

Evaluating this expression at the boundary $r=R$ yields
\begin{equation}
	p_{t}(R)=
	\frac{
		\left[
		r_{0}^{3}(2c_{5}+1)\rho_{S}(c_{5}-1)\kappa
		+(6R-6r_{0})c_{5}^{2}
		+3c_{5}r_{0}
		+3r_{0}
		\right]\pi^{2}
		-3R^{3}(2c_{5}+1)\rho_{S}(c_{5}-1)\kappa
	}
	{6R^{3}\pi^{2}\kappa}.
\end{equation}
The condition $p_{t}(R)=0$ determines the density parameter as
\begin{equation}
	\rho_{S}=
	\frac{
		6\pi^{2}c_{5}^{2}(r_{0}-R)
		-3\pi^{2}r_{0}(c_{5}+1)
	}
	{
		2\pi^{2}c_{5}^{2}\kappa r_{0}^{3}
		-\pi^{2}c_{5}\kappa r_{0}^{3}
		-\pi^{2}\kappa r_{0}^{3}
		-6R^{3}c_{5}^{2}\kappa
		+3R^{3}c_{5}\kappa
		+3R^{3}\kappa
	}.
	\label{rhoC}
\end{equation}

For the radial pressure we obtain
\begin{equation}
	p_{r}(r)=
	\frac{1}{r^{3}\kappa}
	\left[
	2c_{5}r
	-b(r)(1+2c_{5})
	\right],
\end{equation}
which at the boundary reduces to
\begin{equation}
	p_{r}(R)=
	\frac{1}{R^{3}\kappa}
	\left[
	2c_{5}R
	-b(R)(1+2c_{5})
	\right].
\end{equation}
Imposing $p_{r}(R)=0$ leads to
\begin{equation}
	p_{r}(R)=
	-\frac{2c_{5}}{R^{2}\kappa(c_{5}-1)}=0
	\qquad \Longleftrightarrow \qquad
	c_{5}=0,
\end{equation}
namely to a constant redshift function.

However, inserting $c_{5}=0$ into Eq.~\eqref{rhoC} gives
\begin{equation}
	\rho_{S}=
	\frac{3\pi^{2}r_{0}}
	{\pi^{2}\kappa r_{0}^{3}-3R^{3}\kappa}
	<0,
\end{equation}
which is physically unacceptable. Consequently,
this third proposal must also be discarded.

A general remark concerning the analytic profiles \eqref{Phi(r)A}, \eqref{Phi(r)B}, and \eqref{Phi(r)C} is in order. Although these redshift functions can, in principle, be employed in other physical contexts, they are not bounded functions of $r$. It is therefore necessary to introduce an external cutoff in order to maintain physical control over the solution. Within the present framework, such a cutoff is naturally provided by the finite radius $r=R$, where both $p_{r}(r)$ and $p_{t}(r)$ are required to vanish, thereby defining the boundary of the configuration.

\section{Energy Conditions}

We now analyse the null energy condition (NEC) for the various wormhole
models constructed in the previous sections.  The NEC requires
$\rho(r)+p_r(r) \ge 0$, which in terms of the equation-of-state
parameter $\omega(r)\equiv p_r(r)/\rho(r)$ becomes $\omega(r) \ge -1$.
By construction, the NEC is violated at the throat because the flare-out
condition forces $p_r(r_0) = -1/(\kappa r_0^2)$ (see Sec.~\ref{Sec3}).
However, away from the throat it may be possible to have $\omega(r)\ge-1$,
i.e., regions where the NEC is satisfied.  In the following we examine
the behaviour of $\omega(r)$ for each proposal using dimensionless
variables.

\medskip
\noindent
\textbf{Dimensionless formulation.}
Introduce $x = kr$ and $x_0 = kr_0$, where $k=\pi/R$.  The energy density
of the polytropic $n=1$ profile becomes
\begin{equation}
	\rho = \rho_S \frac{\sin x}{x},
\end{equation}
and the derivative of the shape function is
\begin{equation}
	b' = P x \sin x,
\end{equation}
with the dimensionless parameter
\begin{equation}
	P = \frac{\kappa\rho_S R^2}{\pi^2} = \frac{\kappa\rho_S}{k^2}.
\end{equation}
The ratio $b/r$ simplifies to
\begin{equation}
	\frac{b}{r} = \frac{x_0 + P\bigl(\sin x - x\cos x
		- \sin x_0 + x_0\cos x_0\bigr)}{x}.
\end{equation}
For $x_0 \ll \pi$ (throat much smaller than the halo radius) the terms
involving $x_0$ are negligible, giving the approximation
\begin{equation}
	\frac{b}{r} \approx \frac{x_0}{x} + P\frac{\sin x}{x} - P\cos x .
\end{equation}
The flare-out condition $b'(r_0)<1$ then yields
\begin{equation}
	P x_0 \sin x_0 \le 1 \quad\Longleftrightarrow\quad
	P \le \frac{1}{x_0\sin x_0} \approx \frac{1}{x_0^2}.
\end{equation}
A more restrictive condition comes from requiring $b(R)/R<1$, which
gives $P<1$.

\subsection{Original proposal (constant redshift function)}

\begin{figure}[h]
	\centering
	\includegraphics{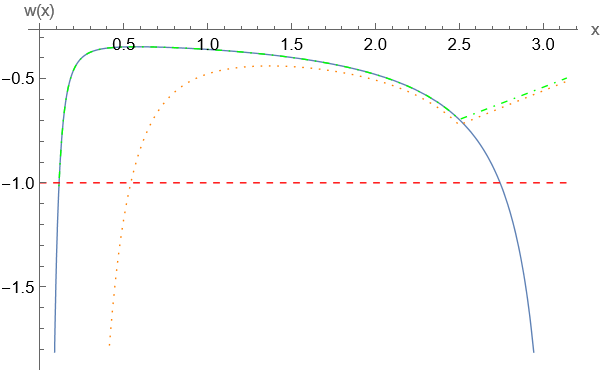}
	\caption{$\omega$ as a function of $x$ for the original proposal
		($\Phi=0$ except for a small cutoff) with $x_0=0.1$.  Solid blue:
		$P=0.5$, no cutoff; dash-dotted green: $P=0.5$, cutoff at
		$\bar{x}=2.5$; dashed orange: $P=0.1$, cutoff at $\bar{x}=2$.
		The horizontal red dashed line marks $\omega=-1$; NEC violation
		occurs below this line.}
	\label{fig1}
\end{figure}

For the model of Sec.~\ref{Sec3} ($\Phi=0$ except in a narrow transition
layer near $r=R$), the equation-of-state parameter is
\begin{equation}
	\omega(x) = -\frac{b}{r b'} = -\frac{1}{x^2}
	+ \frac{\cos x}{x\sin x} - \frac{x_0}{P x^2\sin x}.
\end{equation}
Near $x=\pi$ we introduce a cutoff at $\bar{x}=k\bar{r}$; in that region
the above expression is multiplied by $(\pi-x)/(\pi-\bar{x})$.
Figure~\ref{fig1} shows $\omega(x)$ for $x_0=0.1$, with different values
of $P$ and cutoff positions.  The NEC is violated only in a small
interval near the throat; without the cutoff there would also be a
violation near $x=\pi$.  Decreasing $P$ (i.e., lowering the energy
density) enlarges the NEC-violating region near the throat.

\subsection{Exponent-like redshift function}

\begin{figure}[h]
	\centering
	\includegraphics{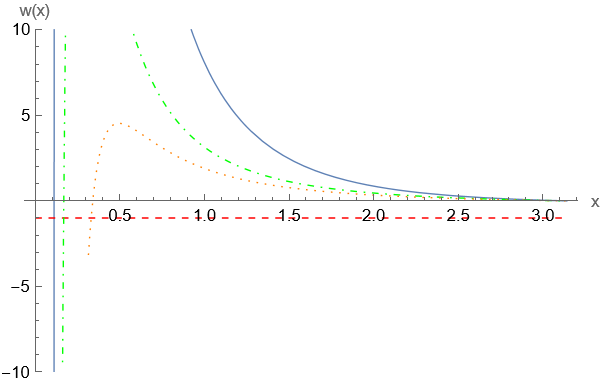}
	\caption{$\omega$ for the exponent-like redshift function,
		$x_0=0.1$.  Solid blue: $P=0.5$; dash-dotted green: $P=0.25$;
		dashed orange: $P=0.1$.  The horizontal red dashed line is
		$\omega=-1$.}
	\label{fig2}
\end{figure}

For the construction of Sec.~\ref{Sec4} with $\Phi'(r)=A f(r)/(2r)$,
$A = b(R)/[(R-b(R))f(R)]$ and $f(r)=K\exp[-r/(R-b(R))]$, we obtain
\begin{equation}
	\omega(x) = \frac{\left(1+\frac{x_0}{\pi P}\right)}{x\sin x}\,
	\frac{1-\frac{x_0}{x}-P\frac{\sin x}{x}+P\cos x}{1-\frac{x_0}{\pi}-P}\,
	\exp\!\left(\frac{-x+\pi}{\pi-P\pi-x_0}\right)
	- \frac{x_0}{P x^2\sin x} + \frac{\cos x}{x\sin x} - \frac{1}{x^2}.
\end{equation}
Figure~\ref{fig2} displays $\omega(x)$ for $x_0=0.1$ and various $P$.
The NEC-violating region is smaller than in the original proposal, and
immediately after it $\omega$ becomes large and positive.

\subsection{Cored-inspired redshift function: Proposals I and II}

\begin{figure}[h]
	\centering
	\includegraphics{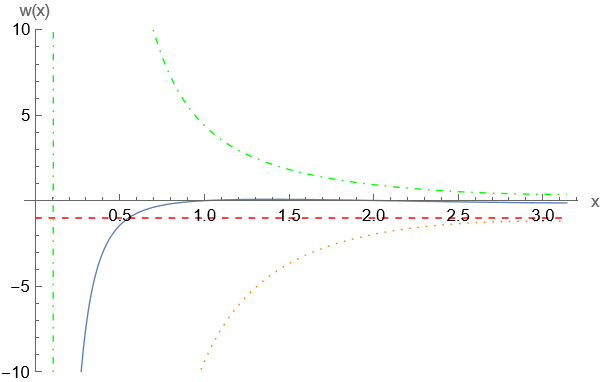}
	\caption{$\omega$ for the cored-inspired redshift function
		($\Phi\propto\sin(x+D)/x$) with $x_0=0.1$, $P=0.5$.  Solid blue:
		$D=0$; dash-dotted green: $D=\pi/4$; dashed orange: $D=\pi/2$.
		The red dashed line is $\omega=-1$.}
	\label{fig3}
\end{figure}

For $\Phi(r)=\Phi_0\sin(kr+D)/(kr)$ (Secs.~\ref{Sec5} and~\ref{Sec6}),
one finds after some algebra
\begin{align}
	\omega(x) = &\frac{\left(1+\frac{x_0}{\pi P}\right)}{x\sin x}\,
	\frac{\cos(x+D)-\frac{\sin(x+D)}{x}}{\frac{\sin D}{\pi}-\cos D}\,
	\frac{1-\frac{x_0}{x}-P\frac{\sin x}{x}+P\cos x}{1-\frac{x_0}{\pi}-P} \\
	&\quad - \frac{x_0}{P x^2\sin x} + \frac{\cos x}{x\sin x} - \frac{1}{x^2}.
\end{align}
The dependence on the phase $D$ is shown in Fig.~\ref{fig3} for
$x_0=0.1$, $P=0.5$.  The size of the NEC-violating region is strongly
influenced by $D$; larger phases can reduce the violation.

\subsection{Cored-inspired redshift function: Proposal III}

\begin{figure}[h]
	\centering
	\includegraphics{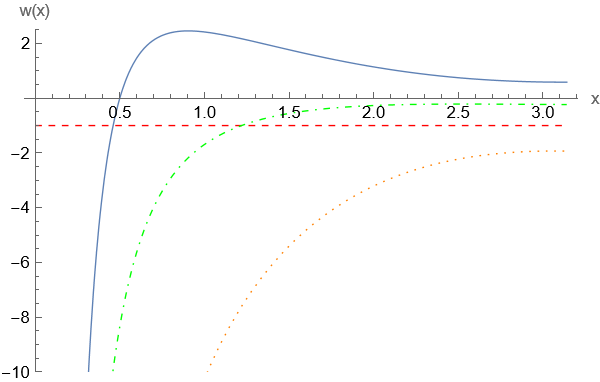}
	\caption{$\omega$ for Proposal III ($\Phi'\propto\sin(x+D)/x$)
		with $x_0=0.1$, $P=0.5$.  Solid blue: $D=7\pi/8$;
		dash-dotted green: $D=\pi/2$; dashed orange: $D=\pi/16$.
		The red dashed line is $\omega=-1$.}
	\label{fig4}
\end{figure}

When $\Phi'(r)$ itself inherits the cored structure,
$\Phi'(r)=\Phi_0\sin(kr+D)/(kr)$ (Sec.~\ref{Sec7}), the equation-of-state
parameter becomes
\begin{equation}
	\omega(x) = -\frac{\left(1+\frac{x_0}{\pi P}\right)}{x\sin x}\,
	\frac{\sin(x+D)}{\sin D}\,
	\frac{1-\frac{x_0}{x}-P\frac{\sin x}{x}+P\cos x}{1-\frac{x_0}{\pi}-P}
	- \frac{x_0}{P x^2\sin x} + \frac{\cos x}{x\sin x} - \frac{1}{x^2}.
\end{equation}
Figure~\ref{fig4} illustrates $\omega(x)$ for $x_0=0.1$, $P=0.5$ and
different $D$.  Larger phases tend to shrink the NEC-violating region.

\subsection{Logarithmic redshift function}

\begin{figure}[h]
	\centering
	\includegraphics{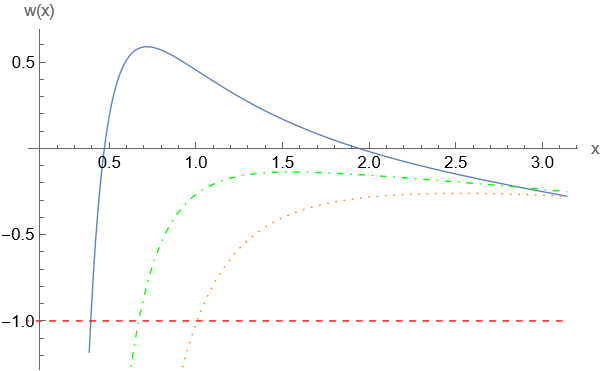}
	\caption{$\omega$ for the logarithmic redshift function
		($x_0=0.1$).  Solid blue: $P=0.5$; dash-dotted green: $P=0.25$;
		dashed orange: $P=0.1$.  The horizontal red dashed line is
		$\omega=-1$.}
	\label{fig5}
\end{figure}

Finally, for the logarithmic redshift function
$\Phi(r)=\ln(c_1 r+c_2)$ (Sec.~\ref{Sec9}), we have $\Phi'=1/(r+\rho)$
with $\rho=c_2/c_1$.  The corresponding $\omega(x)$ reads
\begin{equation}
	\omega(x) = \frac{2\left(1-\frac{x_0}{x}
		- P\frac{\sin x}{x}+P\cos x\right)}{(x+x_\rho)P\sin x}
	- \frac{x_0}{P x^2\sin x} + \frac{\cos x}{x\sin x} - \frac{1}{x^2},
\end{equation}
where $x_\rho$ is chosen to avoid a singularity at $x=\pi$:
\begin{equation}
	x_\rho = \pi\left(\frac{2(1-\frac{x_0}{\pi}-P)}{\frac{x_0}{\pi}+P}-1\right).
\end{equation}
As shown in Fig.~\ref{fig5}, the NEC is again violated only in a narrow
region near the throat.

\medskip
\noindent
\textbf{Summary.}  In all the wormhole models considered, the null
energy condition is violated in a small neighbourhood of the throat,
as required by the flare-out condition.  Away from the throat, however,
it can be satisfied (or even strongly positive) depending on the
parameters.  The size of the NEC-violating region is controlled by
$P$ (related to the energy density of the polytropic halo) and, in the
cored-inspired proposals, by the phase $D$. The logarithmic and
exponent-like redshift functions also show a similar pattern.
Thus, while exotic matter is unavoidable at the throat, its effect can
be confined to a small region, making these wormhole geometries
physically more plausible.
\\

\section{Conclusions and Outlook}\label{Conclusion}

In this work, we have developed a systematic construction of spherically symmetric and static traversable wormholes sustained by exotic matter distributions inspired by dark matter halos. Adopting the polytropic $n=1$ energy density profile as a physically motivated input, we derived the associated shape function $b(r)$ and obtained explicit expressions for the radial and tangential pressures. The geometric requirements at the throat, in particular the flare-out condition and the absence of event horizons, were implemented consistently, leading to well-defined constraints on the free parameters of the model.
A central result of our analysis is that zero-tidal force configurations can be realized through suitably chosen inhomogeneous equations of state. In this setting, the redshift function remains finite throughout the spacetime, ensuring the traversability of the wormhole geometry. Moreover, by extracting candidate redshift profiles directly from the structure of the third Einstein field equation, we demonstrated how analytic ansatze for $\Phi(r)$ can be systematically tested against physically meaningful boundary conditions. This procedure makes explicit the delicate interplay between geometry and matter content, showing that seemingly admissible redshift functions may fail once the full set of constraints is enforced.

Particular attention was devoted to the implementation of boundary conditions at a finite radius $r=R$, interpreted as the edge of the dark matter halo. Requiring the energy density and pressures to vanish smoothly at this boundary not only ensures physical consistency, but also provides a natural infrared cutoff for otherwise unbounded analytic redshift profiles. The asymptotic expansions of the pressures and metric functions near $r=R$ further clarified the behavior of the matter sector in the transition region between the exotic interior and the exterior vacuum spacetime.
An important conceptual outcome of this work is the reformulation of the Einstein field equations entirely in terms of the energy density, radial and tangential pressures, and their derivatives. This matter-driven perspective partially decouples the analysis from the explicit metric functions and yields a flexible framework for exploring a broad class of exotic sources. In particular, it allows one to start from physically motivated density profiles and reconstruct the corresponding geometry in a controlled and systematic way.

The framework developed here naturally suggests several avenues for further investigation. A first extension concerns dynamical wormhole configurations, where time--dependent dark matter profiles could be incorporated to analyze stability, formation mechanisms, and possible evolutionary scenarios. Such a generalization would provide insight into whether these geometries can arise dynamically in realistic astrophysical or cosmological settings.
A second direction involves enriching the physical content of the dark sector. The inclusion of self-interactions, anisotropic stresses, magnetic fields, or rotation could significantly modify both the geometry and the energy conditions, potentially enlarging the class of admissible solutions. In particular, rotating configurations may reveal qualitatively new features, including modified throat structures or frame-dragging effects.

More broadly, the reformulation of the field equations in terms of matter variables opens the possibility of a systematic classification of exotic sources capable of sustaining traversable wormholes. Such a classification could serve as a bridge between purely theoretical constructions and phenomenological models of dark matter. In this perspective, it would be especially interesting to explore possible observational signatures, for instance through gravitational lensing, shadow structures, or perturbative stability analyses, that might distinguish wormhole geometries from black holes in galactic environments.
In summary, the present study provides a coherent and versatile methodology for constructing wormhole solutions supported by dark-matter motivated energy profiles. By clarifying the mutual constraints between geometry, matter content, and boundary conditions, it lays the groundwork for a deeper understanding of exotic compact objects and their potential role in gravitational physics and cosmology.

The wormhole geometries constructed in this work, if they exist within
galactic dark matter halos, may produce distinctive observational
signatures that differ from those of black holes.  Several effects are
worth exploring in future studies.  In gravitational lensing, the
absence of a photon sphere or a different lensing pattern could be
detectable in strong-field observations; for a wormhole with a regular
core, multiple images and relativistic Einstein rings may exhibit
characteristic separations that deviate from black hole predictions
\cite{Tsukamoto:2012xs,Teo:1998dp}.  The shadow of a wormhole can be
smaller or larger than that of a black hole of the same mass, and may
show non-circular features due to anisotropic matter distributions
\cite{Nedkova:2013msa,Gyulchev:2018fmd}; for the polytropic $n=1$
profile, the shadow size could be estimated from the null geodesics of
the metric.  Traversable wormholes typically exhibit a characteristic
ringdown signal with echoes due to the presence of the throat, which
can be searched for in gravitational wave data
\cite{Konoplya:2016hmd,Bueno:2017hyj}.  Additionally, the inner edge of
an accretion disk around a wormhole may lie closer to the compact
object than in the black hole case, leading to higher radiation
efficiency and distinct spectral features \cite{Harko:2009xf}.  A
detailed computation of these signatures for the specific Lane-Emden
$n=1$ wormhole is beyond the scope of the present work.  Nevertheless,
the metric functions derived here provide the necessary foundation for
such future investigations.  By establishing a clear connection between
realistic dark matter profiles and exotic wormhole geometries, this
study opens a new avenue toward observationally testing the existence
of traversable wormholes in galactic environments.

\acknowledgments{
	FSNL acknowledges support from the Funda\c{c}\~{a}o para a Ci\^{e}ncia e a Tecnologia (FCT) Scientific Employment Stimulus contract with reference CEECINST/00032/2018, and funding through the research grants UID/04434/2025 and PTDC/FIS-AST/0054/2021.}

\appendix{}

\section{Coefficients related to Section \ref{Sec5}}

\label{A5}In this appendix we report the coefficients related to the
transverse pressure $p_{t}(r)$%
\begin{align}
	A_{1}\left(  r\right)   &  =\left(  \frac{\Phi_{0}^{2}\left(  \sin\!\left(
		kr_{0}\right)  k\kappa\rho_{S}-\cos\!\left(  kr_{0}\right)  k^{2}\kappa
		r_{0}\rho_{S}+\left(  \left(  r-r_{0}\right)  k^{3}\right)  \right)
		-3r\rho_{S}\kappa k\Phi_{0}}{r^{3}k^{3}\kappa}-\frac{3\Phi_{0}^{2}\rho_{S}%
		\sin\!\left(  kr\right)  }{r^{3}k^{3}}\right)  ,\\
	A_{2}\left(  r\right)   &  =-\frac{\left(  4\Phi_{0}^{2}\kappa\rho_{S}\left(
		\sin\!\left(  kr_{0}\right)  -\cos\!\left(  kr_{0}\right)  kr_{0}\right)
		+k\left(  4\left(  \Phi_{0}^{2}\left(  r-r_{0}\right)  +3\Phi_{0}\kappa
		r^{3}\rho_{S}\right)  k^{2}-6\Phi_{0}r\rho_{S}\kappa\right)  \right)
		\sin\!\left(  kr\right)  }{2r^{4}k^{4}\kappa}\nonumber\\
	&  +\frac{3\Phi_{0}\kappa\rho_{S}\left(  \cos\left(  kr_{0}\right)
		kr_{0}-\sin\left(  kr_{0}\right)  \right)  -\left(  \left(  2r-3r_{0}\right)
		\Phi_{0}\,k^{2}+r\rho_{S}\kappa\right)  k}{2r^{3}k^{3}\kappa},\\
	A_{3}\left(  r\right)   &  =\left(  \Phi_{0}^{2}\kappa\rho_{S}\left(
	\sin\!\left(  kr_{0}\right)  -\cos\!\left(  kr_{0}\right)  kr_{0}\right)
	+k\left(  \Phi_{0}^{2}\left(  r-r_{0}\right)  k^{2}-\Phi_{0}\frac{3r\rho
		_{S}\kappa}{2}\right)  \right)  ,\\
	A_{4}\left(  r\right)   &  =\left(  \left(  k^{2}r^{2}-\frac{3}{2}\right)
	\Phi_{0}\kappa\rho_{S}\left(  \sin\left(  kr_{0}\right)  -kr_{0}\cos\left(
	kr_{0}\right)  \right)  \right)  \nonumber\\
	&  +k\left(  \Phi_{0}r^{2}\left(  r-r_{0}\right)  k^{4}+\left(  \frac{1}%
	{2}r^{3}\rho_{S}\kappa-\Phi_{0}\left(  r+\frac{3}{2}r_{0}\right)  \right)
	k^{2}-\frac{r\rho_{S}\kappa}{2}\right)  ,\\
	A_{5}\left(  r\right)   &  =\frac{3\Phi_{0}k\kappa r\rho_{S}+\kappa\rho
		_{S}\left(  \cos\left(  kr_{0}\right)  kr_{0}-\sin\left(  kr_{0}\right)
		\right)  +r_{0}k^{3}}{2k^{3}r^{3}\kappa}.
\end{align}

\section{Coefficients related to Section \ref{Sec6}}

\label{A6}In this appendix we report the coefficients related to the radial
pressure $p_{r}(r)$%
\begin{align}
	B_{1r}\left(  r\right)   &  =\frac{1}{k^{4}r^{4}\kappa}\left(  2\rho_{S}%
	\kappa\left(  \Phi_{1}kr-\Phi_{2}\right)  \sin\!\left(  kr_{0}\right)
	-2kr_{0}\rho_{S}\kappa\left(  \Phi_{1}kr-\Phi_{2}\right)  \cos\!\left(
	kr_{0}\right)  \right.  \nonumber\\
	&  \left.  +2\left(  \Phi_{1}r\left(  r-r_{0}\right)  k^{2}-\Phi_{2}\left(
	r-r_{0}\right)  k+\frac{r^{2}\rho_{S}\kappa}{2}\right)  k^{2}\right)  ,\\
	B_{2r}\left(  r\right)   &  =2kr_{0}\rho_{S}\kappa\left(  \Phi_{2}kr+\Phi
	_{1}\right)  \cos\!\left(  kr_{0}\right)  -2\rho_{S}\kappa\left(  \Phi
	_{2}kr+\Phi_{1}\right)  \sin\!\left(  kr_{0}\right)  \nonumber\\
	&  -2k\left(  \Phi_{2}r\left(  r-r_{0}\right)  k^{3}+\Phi_{1}\left(
	r-r_{0}\right)  k^{2}+\frac{r\rho_{S}\kappa}{2}\right)  ,\\
	B_{3r}\left(  r\right)   &  =\left(  -2\Phi_{2}\,k^{2}r^{2}-4\Phi_{1}%
	kr+2\Phi_{2}\right)  \rho_{S},\\
	B_{4r}\left(  r\right)   &  =\cos\!\left(  kr_{0}\right)  k\kappa r_{0}%
	\rho_{S}+r_{0}k^{3}-\sin\!\left(  kr_{0}\right)  \kappa\rho_{S}-2\Phi
	_{2}\kappa\rho_{S},
\end{align}
and the coefficients related to the transverse pressure $p_{t}(r)$%
\begin{gather}
	B_{1t}\left(  r\right)  =\left(  -2kr_{0}\rho_{S}\kappa\left(  \Phi_{2}%
	^{2}k^{2}r^{2}+\Phi_{1}^{2}\right)  \cos\!\left(  kr_{0}\right)  +2\rho
	_{S}\kappa\left(  \Phi_{2}^{2}k^{2}r^{2}+\Phi_{1}^{2}\right)  \sin\!\left(
	kr_{0}\right)  \right.  \nonumber\\
	\left.  +\left(  2\Phi_{2}^{2}r^{2}\left(  r-r_{0}\right)  k^{4}+2\Phi_{1}%
	^{2}\left(  r-r_{0}\right)  k^{2}-3\Phi_{1}\kappa r\rho_{S}\right)  k\right)
	,\\
	B_{2t}\left(  r\right)  =2\rho_{S}\kappa\left(  \Phi_{1}^{2}k^{2}r^{2}%
	-4\Phi_{1}\Phi_{2}kr+\Phi_{2}^{2}\right)  \left(  \sin\!\left(  kr_{0}\right)
	-kr_{0}\cos\!\left(  kr_{0}\right)  \right)  +2\Phi_{1}^{2}r^{2}\left(
	r-r_{0}\right)  k^{5}\nonumber\\
	+\left(  2\left(  -3\Phi_{1}\,r^{3}\rho_{S}\kappa+\Phi_{2}^{2}r-\Phi_{2}%
	^{2}r_{0}\right)  k+6\Phi_{2}\,r^{2}\rho_{S}\kappa-r\Phi_{2}\left(  3\kappa
	r^{3}\rho_{S}+8\Phi_{1}r-8\Phi_{1}r_{0}\right)  k^{2}\right)  k^{2},\\
	B_{3t}\left(  r\right)  =\left(  \Phi_{2}^{2}k^{3}r^{3}+3\Phi_{1}^{2}%
	kr-2\Phi_{1}\Phi_{2}\right)  \rho_{S},\\
	B_{4t}\left(  r\right)  =-\rho_{S}\left(  2\Phi_{1}\Phi_{2}\,k^{3}r^{3}%
	+3r^{2}\left(  \Phi_{1}^{2}-\Phi_{2}^{2}\right)  k^{2}-6\Phi_{1}\Phi
	_{2}kr+\Phi_{2}^{2}\right)  ,\\
	B_{5t}\left(  r\right)  =4\rho_{S}\kappa\left(  \Phi_{2}kr+\Phi_{1}\right)
	\left(  \Phi_{1}kr-\Phi_{2}\right)  \left(  kr_{0}\cos\!\left(  kr_{0}\right)
	-\sin\!\left(  kr_{0}\right)  \right)  -4k\Phi_{1}\Phi_{2}\,r^{2}\left(
	r-r_{0}\right)  k^{4}\nonumber\\
	+r\left(  \frac{3\Phi_{1}\,r^{3}\rho_{S}\kappa}{4}+r\left(  \Phi_{1}^{2}%
	-\Phi_{2}^{2}\right)  -\Phi_{1}^{2}r_{0}+\Phi_{2}^{2}r_{0}\right)  k^{3}%
	-\frac{3\Phi_{1}k\,r^{2}\rho_{S}\kappa}{2}\nonumber\\
	+2k\left(  2\Phi_{1}r-2\Phi_{1}r_{0}+3\kappa r^{3}\rho_{S}\right)  \Phi
	_{2}\,k^{2}+\frac{3\Phi_{2}r\rho_{S}\kappa}{4},\\
	B_{6t}\left(  r\right)  =\rho_{S}\kappa\left(  2\Phi_{2}\,k^{2}r^{2}+3\Phi
	_{1}kr-3\Phi_{2}\right)  \left(  kr_{0}\cos\!\left(  kr_{0}\right)
	-\sin\!\left(  kr_{0}\right)  \right)  -2\Phi_{2}\,r^{2}\left(  r-r_{0}%
	\right)  k^{5}\nonumber\\
	-k^{3}\left(  2r-3r_{0}\right)  \left(  \Phi_{1}r\,k-\Phi_{2}\right)
	-\rho_{S}\kappa k^{2}r^{2}+8\Phi_{1}\Phi_{2}k\kappa r\rho_{S}-4\Phi_{2}%
	^{2}\kappa\rho_{S},\\
	B_{7t}\left(  r\right)  =kr_{0}\rho_{S}\kappa\left(  2\Phi_{1}\,k^{2}%
	r^{2}-3\Phi_{2}kr-3\Phi_{1}\right)  \cos\!\left(  kr_{0}\right)  -\rho
	_{S}\kappa\left(  2\Phi_{1}\,k^{2}r^{2}-3\Phi_{2}kr-3\Phi_{1}\right)
	\sin\!\left(  kr_{0}\right)  \nonumber\\
	+r\Phi_{2}\left(  2r-3r_{0}\right)  k^{4}+\left(  -\kappa r^{3}\rho_{S}%
	+2\Phi_{1}r-3\Phi_{1}r_{0}\right)  k^{3}-r\rho_{S}\left(  2\Phi_{2}%
	^{2}-1\right)  \kappa k-4\Phi_{1}\Phi_{2}\kappa\rho_{S}-2\Phi_{1}%
	\,r^{2}\left(  r-r_{0}\right)  k^{5},\\
	B_{8t}\left(  r\right)  =-kr_{0}\rho_{S}\kappa\left(  4\Phi_{1}\Phi
	_{2}-kr\right)  \cos\!\left(  kr_{0}\right)  +\rho_{S}\kappa\left(  4\Phi
	_{1}\Phi_{2}-kr\right)  \sin\!\left(  kr_{0}\right)  \nonumber\\
	+\left(  k^{3}rr_{0}+\Phi_{2}\left(  \kappa r^{3}\rho_{S}+4\Phi_{1}\left(
	r-r_{0}\right)  \right)  k^{2}+3\Phi_{1}k\,r^{2}\rho_{S}\kappa-3\Phi_{2}%
	r\rho_{S}\kappa\right)  k.
\end{gather}

\section{Coefficients related to Section \ref{Sec7}}

\label{A7}In this appendix we report the coefficients related to the
transverse pressure $p_{t}(r)$%
\begin{align}
	C_{1t}\left(  r\right)   &  =\left(  2\kappa\rho_{S}\Phi_{3}\left(  \left(
	\sin\!\left(  kr_{0}\right)  -\cos\!\left(  kr_{0}\right)  kr_{0}\right)
	+\left(  2\Phi_{3}\left(  r-r_{0}\right)  k^{2}+\rho_{S}\kappa\right)
	k\right)  \Phi_{3}\right)  ,\\
	C_{2t}\left(  r\right)   &  =\left(  2\kappa\rho_{S}\Phi_{4}^{2}\left(
	\sin\!\left(  kr_{0}\right)  -kr_{0}\cos\!\left(  kr_{0}\right)  \right)
	+\left(  \left(  3r^{2}\rho_{S}\kappa\Phi_{3}+2\Phi_{4}^{2}\left(
	r-r_{0}\right)  \right)  k-r\rho_{S}\kappa\Phi_{4}\right)  k^{2}\right)  ,\\
	C_{3t}\left(  r\right)   &  =\left(  \left(  4\kappa\rho_{S}\Phi_{3}\Phi
	_{4}\left(  \sin\!\left(  kr_{0}\right)  -\cos\!\left(  kr_{0}\right)
	kr_{0}\right)  -\left(  \Phi_{4}\left(  3r^{2}\rho_{S}\kappa-4\Phi_{3}\left(
	r-r_{0}\right)  \right)  k^{2}+3kr\rho_{S}\kappa\Phi_{3}-\rho_{S}\kappa
	\Phi_{4}\right)  k\right)  \cos\!\left(  kr\right)  \right)  ,\\
	C_{4t}\left(  r\right)   &  =\left(  -\rho_{S}\kappa\left(  2kr\Phi_{4}%
	+\Phi_{3}\right)  \left(  \sin\!\left(  kr_{0}\right)  -kr_{0}\cos\!\left(
	kr_{0}\right)  \right)  +\left(  2r\Phi_{4}\left(  r-r_{0}\right)
	k^{3}+\left(  r^{2}\rho_{S}\kappa-r_{0}\Phi_{3}\right)  k^{2}-\rho_{S}%
	\kappa\right)  k\right)  ,\\
	C_{5t}\left(  r\right)   &  =\left(  -\kappa r_{0}\rho_{S}k\left(  2kr\Phi
	_{3}-\Phi_{4}\right)  \cos\!\left(  kr_{0}\right)  +\rho_{S}\kappa\left(
	2kr\Phi_{3}-\Phi_{4}\right)  \sin\!\left(  kr_{0}\right)  +\left(  2r\Phi
	_{3}\left(  r-r_{0}\right)  k^{2}+kr_{0}\Phi_{4}-r\rho_{S}\kappa\right)
	k^{2}\right)  .
\end{align}

\section{Coefficients related to Section \ref{Sec8}}

\label{A8}In this appendix we report the coefficients related to the
transverse pressure $p_{t}(r)$%
\begin{align}
	D_{1t}\left(  r\right)   &  =4\rho_{S}\kappa rr_{0}\Phi_{5}\Phi_{6}%
	\sin\!\left(  kr_{0}\right)  +2\Phi_{5}^{2}\left(  r_{0}^{2}\rho_{S}\kappa
	\sin\!\left(  kr_{0}\right)  +r_{0}^{2}kr_{0}\left(  k^{2}\left(
	r-r_{0}\right)  -\cos\!\left(  kr_{0}\right)  \rho_{S}\kappa\right)  \right)
	+\Phi_{5}kr_{0}^{2}\kappa\rho_{S},\\
	D_{2t}\left(  r\right)   &  =\Phi_{6}\left[  \kappa rkr_{0}\rho_{S}\left(
	1+k^{2}r^{2}\right)  \sin\!\left(  kr_{0}\right)  -2\left(  2rr_{0}\rho
	_{S}\kappa\sin^{2}\!\left(  kr_{0}\right)  +2rr_{0}\left(  k^{2}\left(
	r-r_{0}\right)  -\cos\!\left(  kr_{0}\right)  r_{0}\rho_{S}\kappa\right)
	k\sin\!\left(  kr_{0}\right)  \right)  \Phi_{5}\right]  ,\\
	D_{3t}\left(  r\right)   &  =\left(  \left(  \cos\left(  kr_{0}\right)
	r_{0}\rho_{S}\kappa+k^{2}r_{0}\right)  r_{0}^{2}k^{2}-r_{0}^{2}k\kappa\rho
	_{S}\sin\left(  kr_{0}\right)  \right)  \Phi_{5}+\kappa k^{2}r_{0}^{2}\rho
	_{S}\left(  1-k^{2}r^{2}\right)  ,\\
	D_{4t}\left(  r\right)   &  =2r^{3}\rho\kappa\Phi_{6}^{2}k\sin^{2}\!\left(
	kr_{0}\right)  -\kappa\rho r\,k^{3}r_{0}^{2}-r_{0}r^{2}\rho\kappa k^{2}%
	\sin\!\left(  kr_{0}\right)  \Phi_{6}\nonumber\\
	&  +\Phi_{5}\left(  2k^{2}r_{0}^{2}r\rho\kappa\sin\left(  kr_{0}\right)
	+2\left(  \left(  r-r_{0}\right)  k^{2}-\cos\!\left(  kr_{0}\right)  r_{0}%
	\rho\kappa\right)  r\,k^{3}r_{0}^{2}\right)  ,\\
	D_{5t}\left(  r\right)   &  =\left(  \sin^{3}\left(  kr_{0}\right)  \kappa
	r^{2}\rho_{S}+\left(  k^{2}r\left(  r-r_{0}\right)  -\cos\!\left(
	kr_{0}\right)  rr_{0}\rho_{S}\kappa\right)  kr\sin^{2}\left(  kr_{0}\right)
	\right)  \Phi_{6}^{2},\\
	D_{6t}\left(  r\right)   &  =\left(  -k\kappa rr_{0}\rho_{S}\sin^{2}\left(
	kr_{0}\right)  +r_{0}\left(  \cos\left(  kr_{0}\right)  rr_{0}\rho_{S}%
	\kappa-r\left(  2r-r_{0}\right)  k^{2}\right)  k^{2}\sin\!\left(
	kr_{0}\right)  \right)  \Phi_{6}.
\end{align}


\begin{thebibliography}{9}
	
	\bibitem{Morris:1988cz}
	M.~S.~Morris and K.~S.~Thorne,
	``Wormholes in space-time and their use for interstellar travel: A tool for teaching general relativity,''
	Am. J. Phys. \textbf{56} (1988), 395-412.
	
	\bibitem{Visser:1995cc}
	M.~Visser,
	``Lorentzian wormholes: From Einstein to Hawking,''
	
	\bibitem{Visser:1989kh}
	M.~Visser,
	``Traversable wormholes: Some simple examples,''
	Phys. Rev. D \textbf{39} (1989), 3182-3184
	[arXiv:0809.0907 [gr-qc]].
	
	\bibitem{Visser:1989kg}
	M.~Visser,
	``Traversable wormholes from surgically modified Schwarzschild space-times,''
	Nucl. Phys. B \textbf{328} (1989), 203-212
	[arXiv:0809.0927 [gr-qc]].
	
	\bibitem{Visser:1999de}
	M.~Visser and C.~Barcelo,
	``Energy conditions and their cosmological implications,''
	[arXiv:gr-qc/0001099 [gr-qc]].
	
	\bibitem{Garattini:2007ff}
	R.~Garattini and F.~S.~N.~Lobo,
	``Self sustained phantom wormholes in semi-classical gravity,''
	Class. Quant. Grav. \textbf{24} (2007), 2401-2413
	[arXiv:gr-qc/0701020 [gr-qc]].
	
	\bibitem{Lobo:2003xd}
	F.~S.~N.~Lobo and P.~Crawford,
	``Linearized stability analysis of thin shell wormholes with a cosmological constant,''
	Class. Quant. Grav. \textbf{21} (2004), 391-404
	[arXiv:gr-qc/0311002 [gr-qc]].
	
	\bibitem{Lobo:2004rp}
	F.~S.~N.~Lobo,
	``Energy conditions, traversable wormholes and dust shells,''
	Gen. Rel. Grav. \textbf{37} (2005), 2023-2038
	[arXiv:gr-qc/0410087 [gr-qc]].
	
	\bibitem{Lobo:2004id}
	F.~S.~N.~Lobo,
	``Surface stresses on a thin shell surrounding a traversable wormhole,''
	Class. Quant. Grav. \textbf{21} (2004), 4811-4832
	[arXiv:gr-qc/0409018 [gr-qc]].
	
	\bibitem{Visser:2003yf}
	M.~Visser, S.~Kar and N.~Dadhich,
	``Traversable wormholes with arbitrarily small energy condition violations,''
	Phys. Rev. Lett. \textbf{90} (2003), 201102
	[arXiv:gr-qc/0301003 [gr-qc]].
	
	\bibitem{Hochberg:1997wp}
	D.~Hochberg and M.~Visser,
	``Geometric structure of the generic static traversable wormhole throat,''
	Phys. Rev. D \textbf{56} (1997), 4745-4755
	[arXiv:gr-qc/9704082 [gr-qc]].
	
	\bibitem{Hochberg:1998ii}
	D.~Hochberg and M.~Visser,
	``The Null energy condition in dynamic wormholes,''
	Phys. Rev. Lett. \textbf{81} (1998), 746-749
	[arXiv:gr-qc/9802048 [gr-qc]].
	
	\bibitem{Flanagan:1996gw}
	E.~E.~Flanagan and R.~M.~Wald,
	``Does back reaction enforce the averaged null energy condition in semiclassical gravity?,''
	Phys. Rev. D \textbf{54} (1996), 6233-6283
	[arXiv:gr-qc/9602052 [gr-qc]].
	
	\bibitem{Ford:1995wg}
	L.~H.~Ford and T.~A.~Roman,
	``Quantum field theory constrains traversable wormhole geometries,''
	Phys. Rev. D \textbf{53} (1996), 5496-5507
	[arXiv:gr-qc/9510071 [gr-qc]].
	
	\bibitem{Lobo:2005us}
	F.~S.~N.~Lobo,
	``Phantom energy traversable wormholes,''
	Phys. Rev. D \textbf{71} (2005), 084011
	[arXiv:gr-qc/0502099 [gr-qc]].
	
	\bibitem{Lobo:2005yv}
	F.~S.~N.~Lobo,
	``Stability of phantom wormholes,''
	Phys. Rev. D \textbf{71} (2005), 124022
	[arXiv:gr-qc/0506001 [gr-qc]].
	
	\bibitem{Lobo:2005vc}
	F.~S.~N.~Lobo,
	``Chaplygin traversable wormholes,''
	Phys. Rev. D \textbf{73} (2006), 064028
	[arXiv:gr-qc/0511003 [gr-qc]].
	
	\bibitem{Lobo:2007zb}
	F.~S.~N.~Lobo,
	``Exotic solutions in General Relativity: Traversable wormholes and 'warp drive' spacetimes,''
	[arXiv:0710.4474 [gr-qc]].
	
	\bibitem{Lobo:2007qi}
	F.~S.~N.~Lobo,
	``A General class of braneworld wormholes,''
	Phys. Rev. D \textbf{75} (2007), 064027
	[arXiv:gr-qc/0701133 [gr-qc]].
	
	\bibitem{Lobo:2008zu}
	F.~S.~N.~Lobo,
	``General class of wormhole geometries in conformal Weyl gravity,''
	Class. Quant. Grav. \textbf{25} (2008), 175006
	[arXiv:0801.4401 [gr-qc]].
	
	\bibitem{Lobo:2006ue}
	F.~S.~N.~Lobo,
	``Van der Waals quintessence stars,''
	Phys. Rev. D \textbf{75} (2007), 024023
	[arXiv:gr-qc/0610118 [gr-qc]].
	
	\bibitem{Dai:2020rnc}
	D.~C.~Dai, D.~Minic and D.~Stojkovic,
	``How to form a wormhole,''
	Eur. Phys. J. C \textbf{80} (2020) no.12, 1103
	[arXiv:2010.03947 [gr-qc]].
	
	\bibitem{Lemos:2003jb}
	J.~P.~S.~Lemos, F.~S.~N.~Lobo and S.~Quinet de Oliveira,
	``Morris-Thorne wormholes with a cosmological constant,''
	Phys. Rev. D \textbf{68} (2003), 064004
	[arXiv:gr-qc/0302049 [gr-qc]].
	
	\bibitem{Lemos:2004vs}
	J.~P.~S.~Lemos and F.~S.~N.~Lobo,
	``Plane symmetric traversable wormholes in an Anti-de Sitter background,''
	Phys. Rev. D \textbf{69} (2004), 104007
	[arXiv:gr-qc/0402099 [gr-qc]].
	
	\bibitem{Nandi:1997mx}
	K.~K.~Nandi, A.~Islam and J.~Evans,
	``Brans wormholes,''
	Phys. Rev. D \textbf{55} (1997), 2497-2500
	[arXiv:0906.0436 [gr-qc]].
	
	\bibitem{Dai:2018vrw}
	D.~C.~Dai, D.~Minic and D.~Stojkovic,
	``New wormhole solution in de Sitter space,''
	Phys. Rev. D \textbf{98} (2018) no.12, 124026
	[arXiv:1810.03432 [hep-th]].
	
	\bibitem{Gao:2016bin}
	P.~Gao, D.~L.~Jafferis and A.~C.~Wall,
	``Traversable Wormholes via a Double Trace Deformation,''
	JHEP \textbf{12} (2017), 151
	[arXiv:1608.05687 [hep-th]].
	
	\bibitem{Maldacena:2017axo}
	J.~Maldacena, D.~Stanford and Z.~Yang,
	``Diving into traversable wormholes,''
	Fortsch. Phys. \textbf{65} (2017) no.5, 1700034
	[arXiv:1704.05333 [hep-th]].
	
	\bibitem{Maldacena:2018gjk}
	J.~Maldacena, A.~Milekhin and F.~Popov,
	``Traversable wormholes in four dimensions,''
	Class. Quant. Grav. \textbf{40} (2023) no.15, 155016
	[arXiv:1807.04726 [hep-th]].
	
	\bibitem{Maldacena:2020sxe}
	J.~Maldacena and A.~Milekhin,
	``Humanly traversable wormholes,''
	Phys. Rev. D \textbf{103} (2021) no.6, 066007
	[arXiv:2008.06618 [hep-th]].
	
	\bibitem{Jensen:2013ora}
	K.~Jensen and A.~Karch,
	``Holographic Dual of an Einstein-Podolsky-Rosen Pair has a Wormhole,''
	Phys. Rev. Lett. \textbf{111} (2013) no.21, 211602
	[arXiv:1307.1132 [hep-th]].
	
	\bibitem{Wall:2010jtc}
	A.~C.~Wall,
	``The Generalized Second Law implies a Quantum Singularity Theorem,''
	Class. Quant. Grav. \textbf{30} (2013), 165003
	[arXiv:1010.5513 [gr-qc]].
	
	\bibitem{Nezami:2021yaq}
	S.~Nezami, H.~W.~Lin, A.~R.~Brown, H.~Gharibyan, S.~Leichenauer, G.~Salton, L.~Susskind, B.~Swingle and M.~Walter,
	``Quantum Gravity in the Lab. II. Teleportation by Size and Traversable Wormholes,''
	PRX Quantum \textbf{4} (2023) no.1, 010321
	[arXiv:2102.01064 [quant-ph]].
	
	\bibitem{Bintanja:2021xfs}
	S.~Bintanja, R.~Esp{\'\i}ndola, B.~Freivogel and D.~Nikolakopoulou,
	``How to make traversable wormholes: eternal AdS$_{4}$ wormholes from coupled CFT{\textquoteright}s,''
	JHEP \textbf{10} (2021), 173
	[arXiv:2102.06628 [hep-th]].
	
	\bibitem{Espindola:2025ons}
	R.~Esp{\'\i}ndola, V.~Jahnke and K.~Y.~Kim,
	``Islands and traversable wormholes,''
	[arXiv:2510.21985 [hep-th]].
	
	\bibitem{Aguilar-Gutierrez:2023ymx}
	S.~E.~Aguilar-Gutierrez, R.~Esp{\'\i}ndola and E.~K.~Morvan-Benhaim,
	``A teleportation protocol in Schwarzschild-de Sitter space,''
	JHEP \textbf{03} (2025), 095
	[arXiv:2308.13516 [hep-th]].
	
	\bibitem{Klinkhammer:1991ki}
	G.~Klinkhammer,
	``Averaged energy conditions for free scalar fields in flat space-times,''
	Phys. Rev. D \textbf{43} (1991), 2542-2548.
	
	\bibitem{Barcelo:1999hq}
	C.~Barcelo and M.~Visser,
	``Traversable wormholes from massless conformally coupled scalar fields,''
	Phys. Lett. B \textbf{466} (1999), 127-134
	[arXiv:gr-qc/9908029 [gr-qc]].
	
	\bibitem{Barcelo:2000zf}
	C.~Barcelo and M.~Visser,
	``Scalar fields, energy conditions, and traversable wormholes,''
	Class. Quant. Grav. \textbf{17} (2000), 3843-3864
	[arXiv:gr-qc/0003025 [gr-qc]].
	
	\bibitem{Barcelo:2000ta}
	C.~Barcelo and M.~Visser,
	``Brane surgery: Energy conditions, traversable wormholes, and voids,''
	Nucl. Phys. B \textbf{584} (2000), 415-435
	[arXiv:hep-th/0004022 [hep-th]].
	
	\cite{Garattini:2008xz}
	\bibitem{Garattini:2008xz}
	R.~Garattini and F.~S.~N.~Lobo,
	``Self-sustained traversable wormholes in noncommutative geometry,''
	Phys. Lett. B \textbf{671} (2009), 146-152
	[arXiv:0811.0919 [gr-qc]].
	
	\bibitem{Bueno:2017hyj}
	P.~Bueno, P.~A.~Cano, F.~Goelen, T.~Hertog and B.~Vercnocke,
	``Echoes of Kerr-like wormholes,''
	Phys. Rev. D \textbf{97} (2018) no.2, 024040
	[arXiv:1711.00391 [gr-qc]].
	
	\bibitem{Bolokhov:2012kn}
	S.~V.~Bolokhov, K.~A.~Bronnikov and M.~V.~Skvortsova,
	``Magnetic black universes and wormholes with a phantom scalar,''
	Class. Quant. Grav. \textbf{29} (2012), 245006
	[arXiv:1208.4619 [gr-qc]].
	
	\bibitem{Kanti:2011jz}
	P.~Kanti, B.~Kleihaus and J.~Kunz,
	``Wormholes in Dilatonic Einstein-Gauss-Bonnet Theory,''
	Phys. Rev. Lett. \textbf{107} (2011), 271101
	[arXiv:1108.3003 [gr-qc]].
	
	\bibitem{Bronnikov:2005gm}
	K.~A.~Bronnikov and J.~C.~Fabris,
	``Regular phantom black holes,''
	Phys. Rev. Lett. \textbf{96} (2006), 251101
	[arXiv:gr-qc/0511109 [gr-qc]].
	
	\bibitem{Bronnikov:2015pha}
	K.~A.~Bronnikov and A.~M.~Galiakhmetov,
	``Wormholes without exotic matter in Einstein{\textendash}Cartan theory,''
	Grav. Cosmol. \textbf{21} (2015) no.4, 283-288
	[arXiv:1508.01114 [gr-qc]].
	
	\bibitem{Bronnikov:2021uta}
	K.~A.~Bronnikov and R.~K.~Walia,
	``Field sources for Simpson-Visser spacetimes,''
	Phys. Rev. D \textbf{105} (2022) no.4, 044039
	[arXiv:2112.13198 [gr-qc]].
	
	\bibitem{Bronnikov:2021liv}
	K.~A.~Bronnikov, R.~A.~Konoplya and T.~D.~Pappas,
	``General parametrization of wormhole spacetimes and its application to shadows and quasinormal modes,''
	Phys. Rev. D \textbf{103} (2021) no.12, 124062
	[arXiv:2102.10679 [gr-qc]].
	
	\bibitem{Lobo:2009ip}
	F.~S.~N.~Lobo and M.~A.~Oliveira,
	``Wormhole geometries in f(R) modified theories of gravity,''
	Phys. Rev. D \textbf{80} (2009), 104012
	[arXiv:0909.5539 [gr-qc]].
	
	\bibitem{Harko:2013yb}
	T.~Harko, F.~S.~N.~Lobo, M.~K.~Mak and S.~V.~Sushkov,
	``Modified-gravity wormholes without exotic matter,''
	Phys. Rev. D \textbf{87} (2013) no.6, 067504
	[arXiv:1301.6878 [gr-qc]].
	
	\bibitem{Boehmer:2012uyw}
	C.~G.~Boehmer, T.~Harko and F.~S.~N.~Lobo,
	``Wormhole geometries in modified teleparralel gravity and the energy conditions,''
	Phys. Rev. D \textbf{85} (2012), 044033
	[arXiv:1110.5756 [gr-qc]].
	
	\bibitem{Mehdizadeh:2015jra}
	M.~R.~Mehdizadeh, M.~Kord Zangeneh and F.~S.~N.~Lobo,
	``Einstein-Gauss-Bonnet traversable wormholes satisfying the weak energy condition,''
	Phys. Rev. D \textbf{91} (2015) no.8, 084004
	[arXiv:1501.04773 [gr-qc]].
	
	\bibitem{Myrzakulov:2015kda}
	R.~Myrzakulov, L.~Sebastiani, S.~Vagnozzi and S.~Zerbini,
	``Static spherically symmetric solutions in mimetic gravity: rotation curves and wormholes,''
	Class. Quant. Grav. \textbf{33} (2016) no.12, 125005
	[arXiv:1510.02284 [gr-qc]].
	
	\bibitem{KordZangeneh:2015dks}
	M.~Kord Zangeneh, F.~S.~N.~Lobo and M.~H.~Dehghani,
	``Traversable wormholes satisfying the weak energy condition in third-order Lovelock gravity,''
	Phys. Rev. D \textbf{92} (2015) no.12, 124049
	[arXiv:1510.07089 [gr-qc]].
	
	\bibitem{Moradpour:2016ubd}
	H.~Moradpour, N.~Sadeghnezhad and S.~H.~Hendi,
	``Traversable asymptotically flat wormholes in Rastall gravity,''
	Can. J. Phys. \textbf{95} (2017) no.12, 1257-1266
	[arXiv:1606.00846 [gr-qc]].
	
	\bibitem{Jamil:2010ziq}
	M.~Jamil, P.~K.~F.~Kuhfittig, F.~Rahaman and S.~A.~Rakib,
	``Wormholes supported by polytropic phantom energy,''
	Eur. Phys. J. C \textbf{67} (2010), 513-520
	[arXiv:0906.2142 [gr-qc]].
	
	\bibitem{Kar:2004hc}
	S.~Kar, N.~Dadhich and M.~Visser,
	``Quantifying energy condition violations in traversable wormholes,''
	Pramana \textbf{63} (2004), 859-864
	[arXiv:gr-qc/0405103 [gr-qc]].
	
	\bibitem{Garcia:2011aa}
	N.~M.~Garcia, F.~S.~N.~Lobo and M.~Visser,
	``Generic spherically symmetric dynamic thin-shell traversable wormholes in standard general relativity,''
	Phys. Rev. D \textbf{86} (2012), 044026
	[arXiv:1112.2057 [gr-qc]].
	
	\bibitem{Mustafa:2021ykn}
	G.~Mustafa, Z.~Hassan, P.~H.~R.~S.~Moraes and P.~K.~Sahoo,
	``Wormhole solutions in symmetric teleparallel gravity,''
	Phys. Lett. B \textbf{821} (2021), 136612
	[arXiv:2108.01446 [gr-qc]].
	
	\bibitem{Boehmer:2007rm}
	C.~G.~Boehmer, T.~Harko and F.~S.~N.~Lobo,
	``Conformally symmetric traversable wormholes,''
	Phys. Rev. D \textbf{76} (2007), 084014
	[arXiv:0708.1537 [gr-qc]].
	
	\bibitem{Brown:2019hmk}
	A.~R.~Brown, H.~Gharibyan, S.~Leichenauer, H.~W.~Lin, S.~Nezami, G.~Salton, L.~Susskind, B.~Swingle and M.~Walter,
	``Quantum Gravity in the Lab. I. Teleportation by Size and Traversable Wormholes,''
	PRX Quantum \textbf{4} (2023) no.1, 010320
	[arXiv:1911.06314 [quant-ph]].
	
	\bibitem{Blok:2020may}
	M.~S.~Blok, V.~V.~Ramasesh, T.~Schuster, K.~O'Brien, J.~M.~Kreikebaum, D.~Dahlen, A.~Morvan, B.~Yoshida, N.~Y.~Yao and I.~Siddiqi,
	``Quantum Information Scrambling on a Superconducting Qutrit Processor,''
	Phys. Rev. X \textbf{11} (2021) no.2, 021010
	[arXiv:2003.03307 [quant-ph]].
	
	\bibitem{Jusufi:2020rpw}
	K.~Jusufi, P.~Channuie and M.~Jamil,
	``Traversable Wormholes Supported by GUP Corrected Casimir Energy,''
	Eur. Phys. J. C \textbf{80} (2020) no.2, 127
	[arXiv:2002.01341 [gr-qc]].
	
	\bibitem{Nascimento:2020ime}
	J.~R.~Nascimento, A.~Y.~Petrov, P.~J.~Porfirio and A.~R.~Soares,
	``Gravitational lensing in black-bounce spacetimes,''
	Phys. Rev. D \textbf{102} (2020) no.4, 044021
	[arXiv:2005.13096 [gr-qc]].
	
	\bibitem{Lobo:2020ffi}
	F.~S.~N.~Lobo, M.~E.~Rodrigues, M.~V.~de Sousa Silva, A.~Simpson and M.~Visser,
	``Novel black-bounce spacetimes: wormholes, regularity, energy conditions, and causal structure,''
	Phys. Rev. D \textbf{103} (2021) no.8, 084052
	[arXiv:2009.12057 [gr-qc]].
	
	\bibitem{Capozziello:2012hr}
	S.~Capozziello, T.~Harko, T.~S.~Koivisto, F.~S.~N.~Lobo and G.~J.~Olmo,
	``Wormholes supported by hybrid metric-Palatini gravity,''
	Phys. Rev. D \textbf{86} (2012), 127504
	[arXiv:1209.5862 [gr-qc]].
	
	\bibitem{Boehmer:2007md}
	C.~G.~Boehmer, T.~Harko and F.~S.~N.~Lobo,
	``Wormhole geometries with conformal motions,''
	Class. Quant. Grav. \textbf{25} (2008), 075016
	[arXiv:0711.2424 [gr-qc]].
	
	\bibitem{Teo:1998dp}
	E.~Teo,
	``Rotating traversable wormholes,''
	Phys. Rev. D \textbf{58} (1998), 024014
	[arXiv:gr-qc/9803098 [gr-qc]].
	
	\bibitem{Harko:2009xf}
	T.~Harko, Z.~Kovacs and F.~S.~N.~Lobo,
	``Thin accretion disks in stationary axisymmetric wormhole spacetimes,''
	Phys. Rev. D \textbf{79} (2009), 064001
	[arXiv:0901.3926 [gr-qc]].
	
	\bibitem{Harko:2008vy}
	T.~Harko, Z.~Kovacs and F.~S.~N.~Lobo,
	``Electromagnetic signatures of thin accretion disks in wormhole geometries,''
	Phys. Rev. D \textbf{78} (2008), 084005
	[arXiv:0808.3306 [gr-qc]].
	
	\bibitem{Tsukamoto:2012xs}
	N.~Tsukamoto, T.~Harada and K.~Yajima,
	``Can we distinguish between black holes and wormholes by their Einstein ring systems?,''
	Phys. Rev. D \textbf{86} (2012), 104062
	[arXiv:1207.0047 [gr-qc]].
	
	\bibitem{Tsukamoto:2016zdu}
	N.~Tsukamoto and T.~Harada,
	``Light curves of light rays passing through a wormhole,''
	Phys. Rev. D \textbf{95} (2017) no.2, 024030
	[arXiv:1607.01120 [gr-qc]].
	
	\bibitem{Konoplya:2005et}
	R.~A.~Konoplya and C.~Molina,
	``The Ringing wormholes,''
	Phys. Rev. D \textbf{71} (2005), 124009
	[arXiv:gr-qc/0504139 [gr-qc]].
	
	\bibitem{Konoplya:2016hmd}
	R.~A.~Konoplya and A.~Zhidenko,
	``Wormholes versus black holes: quasinormal ringing at early and late times,''
	JCAP \textbf{12} (2016), 043
	[arXiv:1606.00517 [gr-qc]].
	
	\bibitem{Konoplya:2021hsm}
	R.~A.~Konoplya and A.~Zhidenko,
	``Traversable Wormholes in General Relativity,''
	Phys. Rev. Lett. \textbf{128} (2022) no.9, 091104
	[arXiv:2106.05034 [gr-qc]].
	
	\bibitem{Gyulchev:2018fmd}
	G.~Gyulchev, P.~Nedkova, V.~Tinchev and S.~Yazadjiev,
	``On the shadow of rotating traversable wormholes,''
	Eur. Phys. J. C \textbf{78} (2018) no.7, 544
	[arXiv:1805.11591 [gr-qc]].
	
	\bibitem{Godani:2018blx}
	N.~Godani and G.~C.~Samanta,
	``Traversable Wormholes and Energy Conditions with Two Different Shape Functions in $f(R)$ Gravity,''
	Int. J. Mod. Phys. D \textbf{28} (2018) no.02, 1950039
	[arXiv:1809.00341 [gr-qc]].
	
	\bibitem{DeFalco:2021ksd}
	V.~De Falco, E.~Battista, S.~Capozziello and M.~De Laurentis,
	``Reconstructing wormhole solutions in curvature based Extended Theories of Gravity,''
	Eur. Phys. J. C \textbf{81} (2021) no.2, 157
	[arXiv:2102.01123 [gr-qc]].
	
	\bibitem{Blazquez-Salcedo:2020czn}
	J.~L.~Bl{\'a}zquez-Salcedo, C.~Knoll and E.~Radu,
	``Traversable wormholes in Einstein-Dirac-Maxwell theory,''
	Phys. Rev. Lett. \textbf{126} (2021) no.10, 101102
	[arXiv:2010.07317 [gr-qc]].
	
	\bibitem{Franzin:2021vnj}
	E.~Franzin, S.~Liberati, J.~Mazza, A.~Simpson and M.~Visser,
	``Charged black-bounce spacetimes,''
	JCAP \textbf{07} (2021), 036
	[arXiv:2104.11376 [gr-qc]].
	
	\bibitem{Simpson:2019cer}
	A.~Simpson, P.~Martin-Moruno and M.~Visser,
	``Vaidya spacetimes, black-bounces, and traversable wormholes,''
	Class. Quant. Grav. \textbf{36} (2019) no.14, 145007
	[arXiv:1902.04232 [gr-qc]].
	
	\bibitem{Rosa:2018jwp}
	J.~L.~Rosa, J.~P.~S.~Lemos and F.~S.~N.~Lobo,
	``Wormholes in generalized hybrid metric-Palatini gravity obeying the matter null energy condition everywhere,''
	Phys. Rev. D \textbf{98} (2018) no.6, 064054
	[arXiv:1808.08975 [gr-qc]].
	
	\bibitem{Dai:2019mse}
	D.~C.~Dai and D.~Stojkovic,
	``Observing a Wormhole,''
	Phys. Rev. D \textbf{100} (2019) no.8, 083513
	[arXiv:1910.00429 [gr-qc]].
	
	\bibitem{Simonetti:2020ivl}
	J.~H.~Simonetti, M.~J.~Kavic, D.~Minic, D.~Stojkovic and D.~C.~Dai,
	``Sensitive searches for wormholes,''
	Phys. Rev. D \textbf{104} (2021) no.8, L081502
	[arXiv:2007.12184 [gr-qc]].
	
	\bibitem{Bambi:2021qfo}
	C.~Bambi and D.~Stojkovic,
	``Astrophysical Wormholes,''
	Universe \textbf{7} (2021) no.5, 136
	[arXiv:2105.00881 [gr-qc]].
	
	\bibitem{Bambi:2013nla}
	C.~Bambi,
	``Can the supermassive objects at the centers of galaxies be traversable wormholes? The first test of strong gravity for mm/sub-mm very long baseline interferometry facilities,''
	Phys. Rev. D \textbf{87} (2013), 107501
	[arXiv:1304.5691 [gr-qc]].
	
	\bibitem{Bambi:2013jda}
	C.~Bambi,
	``Broad K{\ensuremath{\alpha}} iron line from accretion disks around traversable wormholes,''
	Phys. Rev. D \textbf{87} (2013), 084039
	[arXiv:1303.0624 [gr-qc]].
	
	\bibitem{Bambi:2015kza}
	C.~Bambi,
	``Testing black hole candidates with electromagnetic radiation,''
	Rev. Mod. Phys. \textbf{89} (2017) no.2, 025001
	[arXiv:1509.03884 [gr-qc]].
	
	\bibitem{Nedkova:2013msa}
	P.~G.~Nedkova, V.~K.~Tinchev and S.~S.~Yazadjiev,
	``Shadow of a rotating traversable wormhole,''
	Phys. Rev. D \textbf{88} (2013) no.12, 124019
	[arXiv:1307.7647 [gr-qc]].
	
	\bibitem{Molina-Paris:1998xmn}
	C.~Molina-Paris and M.~Visser,
	``Minimal conditions for the creation of a Friedman-Robertson-Walker universe from a 'bounce',''
	Phys. Lett. B \textbf{455} (1999), 90-95
	[arXiv:gr-qc/9810023 [gr-qc]].
	
	\bibitem{Boehmer:2007um}
	C.~G.~Boehmer and T.~Harko,
	``Can dark matter be a Bose-Einstein condensate?,''
	JCAP \textbf{06} (2007), 025
	[arXiv:0705.4158 [astro-ph]].
	
	\bibitem{Xu:2020wfm}
	Z.~Xu, M.~Tang, G.~Cao and S.~N.~Zhang,
	``Possibility of traversable wormhole formation in the dark matter halo with istropic pressure,''
	Eur. Phys. J. C \textbf{80} (2020) no.1, 70.
	
	\bibitem{Jusufi:2019knb}
	K.~Jusufi, M.~Jamil and M.~Rizwan,
	``On the possibility of wormhole formation in the galactic halo due to dark matter Bose-Einstein condensates,''
	Gen. Rel. Grav. \textbf{51} (2019) no.8, 102
	[arXiv:1903.01227 [gr-qc]].
	
	\bibitem{Muniz:2022eex}
	C.~R.~Muniz and R.~V.~Maluf,
	``A class of traversable wormholes in the Starobinsky-like f(R) gravity with anisotropic dark matter,''
	Annals Phys. \textbf{446} (2022), 169129.
	
	\bibitem{Ovgun:2018uin}
	A.~Ovg{\"u}n,
	``Evolving topologically deformed wormholes supported in the dark matter halo,''
	Eur. Phys. J. Plus \textbf{136} (2021) no.10, 987
	[arXiv:1803.04256 [physics.gen-ph]].
	
	\bibitem{Kumar:2024vko}
	J.~Kumar, S.~K.~Maurya and S.~Kiroriwal,
	``Dark matter influences on wormhole stability in de Rham{\textendash}Gabadadze{\textendash}Tolley like massive gravity,''
	Eur. Phys. J. C \textbf{84} (2024) no.12, 1305.
	
	\bibitem{Garattini:2021kca}
	R.~Garattini,
	``Yukawa{\textendash}Casimir wormholes,''
	Eur. Phys. J. C \textbf{81} (2021) no.9, 824
	[arXiv:2107.09276 [gr-qc]].
	
	\bibitem{Garattini:2023wgk}
	R.~Garattini and P.~Channuie,
	``Traversable wormholes supported by holographic dark energy with a modified equation of state,''
	Nucl. Phys. B \textbf{1005} (2024), 116589
	[arXiv:2311.04620 [gr-qc]].
	
\end{thebibliography}
\end{document}